\documentclass{aa}
\usepackage[T1]{fontenc}
\usepackage{graphicx}	
\usepackage{amsmath}	
\usepackage{amssymb}	
\usepackage{xcolor}
\usepackage{natbib}
\usepackage{color,array}

\usepackage[normalem]{ulem}  

\newcommand{\respA}[1]{{#1}}
\newcommand{\respB}[1]{{#1}}

\makeatletter
\def\zapcolorreset{\let\reset@color\relax\ignorespaces}
\def\colorrows#1{\noalign{\aftergroup\zapcolorreset#1}\ignorespaces}
\edef\endtabular{\unexpanded\expandafter{\endtabular}\noexpand\reset@color}
\makeatother

\makeatletter
\renewcommand*\aa@pageof{, page \thepage{} of \pageref*{LastPage}}
\makeatother

\begin{document}
\title{Testing kinematic distances under a realistic Galactic potential}

\subtitle{Investigating systematic errors in the kinematic distance method arising from a non-axisymmetric potential}

\author{
Glen H. Hunter \inst{\ref{ITA}}\fnmsep \and
Mattia C. Sormani \inst{\ref{DiSAT},\ref{UoS}} \and
Jan P. Beckmann \inst{\ref{ITA}} \and
Eugene Vasiliev \inst{\ref{IoA},\ref{UoS}} \and
Simon C.\ O.\ Glover  \inst{\ref{ITA}} \and
Ralf S.\ Klessen  \inst{\ref{ITA}, \ref{IWR}, \ref{CfA}, \ref{Rad}} \and 
Juan D. Soler\inst{\ref{IAPS}} \and
No\'e Brucy \inst{\ref{ITA},\ref{ENSL}} \and
Philipp Girichidis  \inst{\ref{ITA}} \and
Junia G\"{o}ller \inst{\ref{ITA}} \and 
Loke Ohlin \inst{\ref{ITA}} \and
Robin Tress \inst{\ref{EPFL}}\and 
Sergio Molinari\inst{\ref{IAPS}}  \and
Ortwin Gerhard \inst{\ref{MPE}} \and
Milena Benedettini \inst{\ref{IAPS}} \and
Rowan Smith \inst{\ref{StA}} \and
Patrick Hennebelle\inst{\ref{CEA}}  \and
Leonardo Testi \inst{\ref{DFA},\ref{OAA}}
}

\institute{
Universit\"{a}t Heidelberg, Zentrum f\"{u}r Astronomie, Institut f\"{u}r Theoretische Astrophysik, Albert-Ueberle-Str. 2, 69120 Heidelberg, Germany \label{ITA}
\and
Universit{\`a} dell’Insubria, via Valleggio 11, 22100 Como, Italy\label{DiSAT}
\and
Department of Physics, University of Surrey, Guildford GU2 7XH, UK \label{UoS}
\and
Institute of Astronomy, University of Cambridge, Madingley Road, Cambridge CB3 0HA, U.K. \label{IoA}
\and
Universit\"{a}t Heidelberg, Interdisziplin\"{a}res Zentrum f\"{u}r Wissenschaftliches Rechnen, Im Neuenheimer Feld 205, 69120 Heidelberg, Germany \label{IWR}
\and 
Harvard-Smithsonian Center for Astrophysics, 60 Garden Street, Cambridge, MA 02138, USA \label{CfA}
\and
Elizabeth S. and Richard M. Cashin Fellow at the Radcliffe Institute for Advanced Studies at Harvard University, 10 Garden Street, Cambridge, MA 02138, USA \label{Rad}
\and
INAF -- Istituto di Astrofisica e Planetologia Spaziali, Via Fosso del Cavaliere 100, 00133 Roma, Italy \label{IAPS}
\and
Centre de Recherche Astrophysique de Lyon UMR5574, ENS de Lyon, Univ. Lyon1, CNRS, Université de Lyon, 69007, Lyon, France \label{ENSL}
\and
Institute of Physics, Laboratory for galaxy evolution and spectral modelling, EPFL, Observatoire de Sauverny, Chemin Pegais 51, 1290 Versoix, Switzerland \label{EPFL}
\and
Max-Planck-Institut f\"{u}r extraterrestrische Physik, Giessenbachstrasse, 85748 Garching, Germany \label{MPE}
\and
School of Physics and Astronomy, University of St. Andrews, North Haugh, St. Andrews, Fife KY16 9SS, UK \label{StA}
\and
Laboratoire AIM, Paris-Saclay, CEA/IRFU/SAp - CNRS - Universit\'{e} Paris Diderot. 91191, Gif-sur-Yvette Cedex, France \label{CEA}
\and
Dipartimento di Fisica e Astronomia “Augusto Righi”, Viale Berti Pichat 6/2, Bologna, Italy \label{DFA}
\and
INAF – Osservatorio Astrofisico di Arcetri, Largo E. Fermi 5, 50125 Firenze, Italy \label{OAA}
}

\date{Accepted XXX. Received YYY; in original form ZZZ}

\abstract{Obtaining reliable distance estimates to gas clouds within the Milky Way is challenging in the absence of certain tracers. The kinematic distance approach has been used as an alternative, derived from the assumption of circular trajectories around the Galactic centre. Consequently, significant errors are expected in regions where gas flow deviates from purely circular motions.
}
{We aim to quantify the systematic errors that arise from the kinematic distance method in the presence of a Galactic potential that is non-axisymmetric. We investigate how these errors differ in certain regions of the Galaxy and how they relate to the underlying dynamics.}
{We perform 2D isothermal hydrodynamical simulation of the gas disk with the moving-mesh code \textsc{Arepo}, adding the capability of using an external potential provided by the \textsc{Agama} library for galactic dynamics. We introduce a new analytic potential of the Milky Way, taking elements from existing models and adjusting parameters to match recent observational constraints.}
{\respB{In line with results of previous studies, we report} significant errors in the kinematic distance estimate for gas close to the Sun, along sight lines towards the Galactic centre and anti-centre, and \respB{associated} with the Galactic bar. Kinematic distance errors are low within the spiral arms as gas resides close to local potential minima and the resulting line-of-sight velocity is \respB{similar} to what is expected for an axisymmetric potential. Interarm regions exhibit large deviations at any given Galactic radius. This is caused by the gas being sped up or slowed down as it travels into or out of spiral arms.  \respB{In addition, we identify} ‘zones of avoidance’ in the $lv$-diagram, where the kinematic distance method is particularly unreliable and should only be used with caution, \respB{and we find} a power law relation between the kinematic distance error and the deviation of the \respA{projected} line-of-sight velocity from circular motion.}{}


\maketitle
\nolinenumbers
\section{Introduction}
\label{sec:intro}
Accurate distance measurements are essential for many fields  of astronomy and astrophysics \cite[e.g.][]{Carroll2017}. Whereas high-precision astrometric data are readily available within the Milky Way for the stellar component (see, e.g.,  \textit{Gaia} data release DR3, \citealt{Gaia2023-DR3}), obtaining reliable distance estimates for the gaseous component, i.e.\ for the various phases of interstellar medium \cite[ISM, see e.g.,][]{Tielens2005, Draine2011} is much more challenging. Estimating distances to molecular clouds is important for understanding their properties, formation and evolution \citep{Molinari2014} and their ability to form stars \citep{KlessenGlover2016}. The same is true for the atomic and ionised components of the ISM. 

Accurate 3D maps of the gas distribution in the Solar Neighbourhood have been constructed by combining precise parallax measurements of stars from \textit{Gaia} and photometric measurements of reddening to the same stars \citep{Lallement2019,Leike2020,Zucker2021}. However, this approach is currently feasible only for a limited volume of a few kpc around the Sun. Reliable distances to clouds further away can be obtained from parallax measurements of molecular maser emission from high mass star-forming regions \citep{Reid2014,Reid2019}, but this approach is time-consuming and cannot be applied to large surveys containing thousands of clouds.

A widely used method to estimate distances to the ISM out to tens of kpc from the Sun is the kinematic distance (KD) method. This method allows one to derive the distance to a molecular cloud from its line-of-sight (LOS) velocity. Historically, it was developed by \cite{vandeHulst1954} and \cite{Oort1958}, who used it to derive the first face-on maps of atomic hydrogen in the Milky Way from 21-cm spectral line observations. The same approach has since been applied several times to produce face-on maps of the neutral and molecular gas in the Milky Way \citep[e.g.][]{Nakanishi2003,Nakanishi2006,Levine2006,Soler2022} and the associated star-formation rate surface density \citep{Elia2022}.

A key assumption of the KD approach is that the gas is in purely circular motion around the Galactic centre. Significant errors in the KD estimations arise if there are deviations from circular motions. \cite{Wenger2018} recently compared kinematic and parallax distances for a sample of 75 Galactic high mass star-forming regions, most of which are at distances $d<10\, \rm kpc$, and found \respA{for their sample} that kinematic distances \respA{were overestimated by $\sim 20$\% on average, with a standard deviation of $\sim 50$\%, when using the KD standard approach with the \citet{Brand1993} rotation model for the Milky Way (their Model A). This overestimation drops below $20$\% when using the updated rotation model of \citet{Reid2014} (see Table 4 of \citealt{Wenger2018} for a complete summary).}

Errors arising from deviations from circular motions can be broadly divided into two categories: (i) random fluctuations around the average streaming motions that do not change the average velocity (e.g., a turbulent velocity dispersion); (ii) systematic changes in the streaming velocity due to non-axisymmetric features such as spiral arms and the Galactic bar. 

\cite{Reid2022} studied the effects of random motions {\respA{in 3D}} on the KD estimates and found that a velocity dispersion of $\sim 7 {\rm \, km \, s^{-1}}$, representative of turbulent motions in giant molecular clouds, can lead to significant ($>10$\%) errors in the KD estimates for true distances $d \lesssim 5 \, \rm kpc$, and can also lead to systematic biases of $\sim 20$\% despite the random motions having zero mean around the underlying circular motions. \respB{A particularly extreme example of deviations from circular velocity leading to errors in the kinematic distance estimate is reported for the Perseus Arm at Galactic longitudes  $\ell \sim 130^\circ$ \cite[see][]{Xu2006}. As pointed out by \cite{Sofue2011} in a detailed study of the physical origins of kinematic distance uncertainties, it is expected that these deviations are more pronounced at small Galactic radii and  longitudes (see also the discussion by \citealt{Sofue2023} for the very inner regions of the Milky Way). }

A number of authors have investigated the effects of streaming motions due to spiral arms using simplified models of the Milky Way. \respB{For example,} \cite{Gomez2006} used 2D hydrodynamical simulations with a simple externally imposed two-armed spiral pattern to compare KD with true distances, and found that errors can be large at the position of the spiral arms. \cite{Baba2009} employed self-consistent $N$-body + hydrodynamical simulations with a live stellar potential, and found that transient and recurring spiral arms can drive strong non-circular motions, \respB{meaning that KD estimates, whilst highly correlated with the true distance, can produce errors of $2-3$ kpc}. Also using hydrodynamical simulations, \cite{Ramon-Fox2018} found that streaming motions can produce systematic offsets of $\sim 1\, \rm kpc$, errors of $\sim 2\, \rm kpc$, and that the results are sensitive to the assumed spiral arm perturbations. Some works have tried to correct the KD method to account for the systematic non-circular motions due to spiral arms and bar, but the results are affected by large uncertainties in the gas streaming motions arising from these components \citep{Foster2006,Pohl2008}. \respB{For a summary of the key geometric and physical concepts behind these various sources of KD uncertainties, we refer to \citet{Sofue2011} and \citet{Wenger2018}.}

The goal of this paper is to \respB{further characterise} the KD uncertainties caused by streaming motions due to spiral arms and the bar as a function of position in the Galaxy using a much more accurate Milky Way model than previous work, and therefore construct maps of the expected systematic uncertainties that can provide useful guidance as to when the KD method should be considered reliable and when it should be avoided. To do this, we construct a realistic model of the Galactic gravitational potential that includes state-of-the-art constraints on the Galactic bar, Galactic disk, dark matter halo and spiral arms, and run 2D hydrodynamical simulations using this potential. We then compare actual and kinematic distances in the model, paying particular attention to the inner regions of the Galaxy dominated by the bar and to the regions around the spiral arms.

After a brief discussion of the context of this study in Section~\ref{sec:intro}, we introduce our new analytic description of the Milky Way potential and its various components in Section~\ref{sec:potential}. We briefly describe our numerical approach and the implementation of the new potential in the \textsc{Agama} frameworks combined with {\sc Arepo} in Section~\ref{sec:methods}. Our main findings are presented in Section~\ref{sec:results}, and their implications and limitations discussed in Section~\ref{sec:discussion}. Finally, we summarise and conclude in Section~\ref{sec:conclusion}.  
\begin{figure}
    \includegraphics{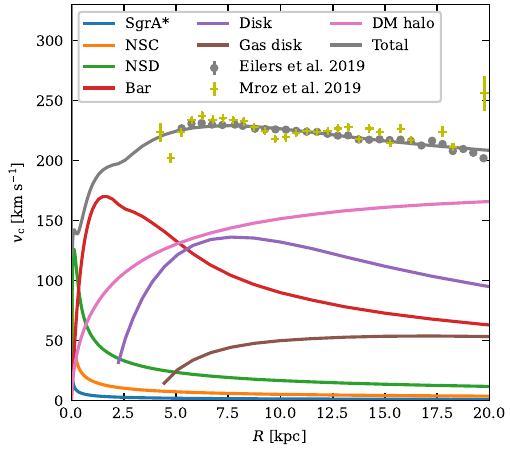}
    \caption{\respB{Circular-velocity curve produced by our model (black line). The contribution of each component of the potential represented by the coloured lines as detailed in the legend. Observation points from \citet{Eilers2019} and \citet{Mroz2019} are also included (colored markers).}
    }
    \label{fig:rot_curve}
\end{figure}

\section{Galactic potential} 
\label{sec:potential}
We introduce a new gravitational potential for our Milky Way-like simulation in order to investigate how non-axisymmetric perturbations affect kinematic distance estimates. The potential comprises of many components, each providing structure in different parts of the Galaxy, as detailed in the following sections.

\begin{figure}
    \includegraphics[width=0.5\textwidth]{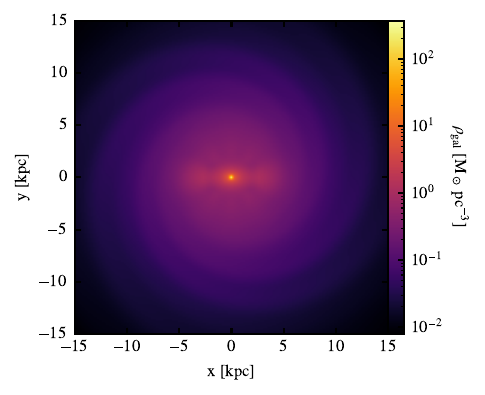}
    \caption{\respA{Underlying density distribution within the midplane ($z=0$ kpc) of the gravitational potential of the Galaxy.}}
    \label{fig:dens}
\end{figure}

The corresponding circular-velocity curves are shown in Figure~\ref{fig:rot_curve}, and the total midplane density profile of our potential, $\rho_\mathrm{gal}$, is shown in Figure~\ref{fig:dens}.

\subsection{Components of the potential}

\subsubsection{Supermassive black hole Sgr~A$^\star$}
\label{sec:SgrA}
The potential of the central supermassive black hole, Sgr A$^\star$, is represented by a \citet{Plummer1911} model:
\begin{equation}\label{eq:plummer}
    \Phi_{\mathrm{Sgr A}^\star} = -\frac{GM_{\mathrm{Sgr A}^\star}}{\sqrt{r^2+b^2}}\;,
\end{equation}
where $G$ is the gravitational constant, the mass $M_{\mathrm{Sgr A}^\star} = 4.154\times 10^6\,\mathrm{M}_\odot$ is taken from the \citet{GRAVITYCollaboration2019}, $r$ is the spherical Galactic radius
and the scale radius $b$ is set to 0.1~pc to avoid a singularity in the potential.

\subsubsection{Nuclear star cluster}
\label{sec:NSC}
The cluster of stars around Sgr~A$^\star$ also contributes to the potential within the core of the Galaxy, dominating in the innermost few pc. In our model, the nuclear star cluster (NSC) follows a flattened \citet{Dehnen1993} density profile as given in \citet[][see their Equation 17]{Chatzopoulos2015}:
\begin{equation} \label{eq:NSC}
    \rho_\mathrm{NSC} = \frac{(3-\gamma)M_\mathrm{NSC}}{4\pi q} \frac{a_0}{a^\gamma(a+a_0)^{4-\gamma}}\;,
\end{equation}
where
\begin{equation} \label{eq:a}
    a(R,z) = \sqrt{R^2+\frac{z^2}{q^2}}\;.
\end{equation}
The parameters $\gamma=0.71$, $q = 0.73$, $a_0 = 5.9$~pc and $M_\mathrm{NSC} = 6.1\times10^7$M$_\odot$ are taken from their best-fitting model. Here $R$ refers to the Galactic radius in cylindrical coordinates. 
We note from our circular-velocity curve (Figure~\ref{fig:rot_curve}) we see little contribution to the overall potential from Sgr~A$^\star$ and the NSC. This is due to these components being most dominant in the inner most 100 pc of the galaxy making it difficult to compare against observational rotation curves and terminal velocities. However, we do include these components for the sake of completeness.

\subsubsection{Nuclear stellar disk}
\label{sec:NSD}
For the nuclear stellar disk (NSD) surrounding the nuclear region, we adopt the parameterisation from the Jeans modelling analysis of \citet{Sormani2020}  
based on data from the APOGEE survey \citep{Majewski2017,Ahumada2020} and the 86~GHz SiO maser survey of \citet{Messineo2002,Messineo2004,Messineo2005}. The density of this component can be written as:
\begin{equation} \label{eq:NSD}
    \rho_\mathrm{NSD} = \rho_1 \exp{\Big[-\Big(\frac{a}{R_1}\Big)^{n_1}\Big]} + \rho_2 \exp{\Big[-\Big(\frac{a}{R_2}\Big)^{n_2}\Big]}
\end{equation}
where $a$ is as defined in Equation~\eqref{eq:a} but with $q=0.37$, and where $n_1=0.72$, $n_2=0.79$, $R_1=5.06$ pc, $R_2=24.6$ pc, $\rho_1/\rho_2=1.311$ and $\rho_2=153\times10^{10}$ M$_\odot$ kpc$^{-3}$, which follows model 3 of \citet{Sormani2020}. As shown in Figure~\ref{fig:rot_curve}, this component dominates between the inner $\sim$20 pc and $\sim$300 pc of the Galaxy. We opt this model over the more recent model of \citet{Sormani2022a} as the density profile of \citet{Sormani2020} is available in a closed, analytical form. This difference only affects the inner most $\sim$300 pc of the simulation which is only a minor impact in comparison to the larger scale of the whole Galaxy.

\subsubsection{Galactic bar}
\label{sec:bar}
The Galactic bar dominates much of the potential within the inner $\sim$5 kpc of the Galaxy. The most realistic model for this component is the made-to-measure (m2m) model from \citet{Portail2017}. It is constrained using red giant stellar density measurements and kinematics from multiple surveys across the entire bar region. Here we make use of the analytical approximation of this model presented by \citet{Sormani2022}, who provide density functions to describe the X-shaped box/peanut bar and the long bar. We reiterate the functions used and their parameters here to have a complete description of our potential in this paper:
\begin{equation} \label{eq:bar}
    \rho_\mathrm{bar} = \underbrace{\rho_{\mathrm{bar},1} + \rho_{\mathrm{bar},2}}_{\text{bar}} + \underbrace{\rho_{\mathrm{bar},3}}_{\text{long bar}} \,.
\end{equation}
The first component of Equation~\eqref{eq:bar} corresponds to the X-shaped component of the observed boxy-peanut shape of the bar \citep{Wegg2013}. To describe this, we use a modified form of Equations 9 and 10 of \citet{Coleman2020} and \citet{Freudenreich1998}, respectively:
\begin{align}
\rho_{\mathrm{bar},1}(x,y,z) & = \rho_{1} \mathrm{sech}\left( a^{m} \right) \nonumber \\
& \times \left[ 1+ \alpha \left(e^{-{a_+^n}}+e^{-{a_-^n}}\right) \right] e^{-\left( \frac{r}{r_{\rm cut}} \right)^2} \,,
\end{align}
where
\begin{align}
a 		& = \left\{ \left[  \left(\frac{|x|}{x_1}\right)^{c_{\perp}} + \left(\frac{|y|}{y_1}\right)^{c_{\perp}} \right]^{\frac{c_\parallel}{c_\perp}} +  \left(\frac{|z|}{z_1}\right)^{c_{\parallel}} \right\}^{\frac{1}{c_\parallel}}\,, \\
a_{\pm}    & =  \left[ \left(\frac{x \pm c z}{x_c}\right)^2  + \left(\frac{y}{y_c}\right)^2  \right]^{\frac{1}{2}} \,, \\
r                & = \left( x^2 + y^2 + z^2 \right)^{\frac{1}{2}} \,.
\end{align}
Here, $\alpha = 0.626$ defines the strength of the X-shape whilst $c=1.342$ defines the slope of the X-shape in the $x-z$ plane. The scale lengths $x_1=0.49$ kpc, $y_1=0.392$ kpc, $z_1=0.229$ kpc, $x_c=0.751$ kpc and $y_c=0.469$ kpc, shape the bar along with the shaping parameters $c_\perp=2.232$ and $c_\parallel=1.991$. The X-shape of the bar trails off with power law exponents with powers $m=0.873$ and $n=1.94$ with an additonal cutoff radius at $r_\mathrm{cut}=4.37$ kpc. The density profile is normalised to $\rho_1 = 3.16\times10^9$ M$_\odot$ kpc$^{-3}$.

The second and third component of $\rho_{\rm bar}$ describe the ellipsoid shape of the bar, which we split into short ($\rho_{2}$) and long ($\rho_{3}$) bar components. Both components follow a modified version of Equation 9 of \citet{Wegg2015} {\respB{with fit parameters determined by \citet{Sormani2022} based on $N$-body simulations of \citet{Portail2017}:}}

\begin{align}
\rho_{\rm bar, i}(x,y,z) & = \rho_i \, e^{-a_i^{n_i}}  \mathrm{sech}^2 \left( \frac{z}{z_i} \right)\nonumber \\
& \times  e^{ - \left(\frac{R}{R_{\rm i, out}} \right)^{n_{\rm i, out}}} e^{- \left(\frac{R_{\rm i,in} }{R} \right)^{n_{\rm i,in}} } \,,
\end{align}
where $i = \{2,3\}$ and
\begin{align}
a_{i} & = \left[  \left(\frac{|x|}{x_{i}}\right)^{c_{\perp,i}} + \left(\frac{|y|}{y_i}\right)^{c_{\perp,i}} \right]^{\frac{1}{c_{\perp,i}}} \,, \\
R      & =  \left( x^2 + y^z \right)^{\frac{1}{2}}  \label{eq:R} \,.
\end{align}

We summarise the parameters used for the components 2 and 3 of the bar in Table~\ref{tab:bar_param}. The total mass contained within the bar is $M_\mathrm{bar}=1.83\times10^{10}$ M$_\odot$.

\begin{table}
    \centering
    \caption{Parameters for components 2 and 3 of  the bar. \label{tab:bar_param}}
    \begin{tabular}{lcc}
        \hline
        Parameter &  Value \\\hline
        & Component 2 & Component 3\\\hline
        $\rho_i$ [M$_\odot$ kpc$^{-3}$] & $0.5\times10^9$ & $1.743\times10^{13}$ \\
        $x_i$ [kpc] & $5.364$ & $0.478$\\
        $y_i$ [kpc] & $0.959$ & $0.297$\\
        $z_i$ [kpc] & $0.611$ & $0.252$\\
        $R_{i,{\rm in}}$ [kpc] & $0.558$ & $7.607$\\
        $R_{i,{\rm out}}$ [kpc] & $3.19$ & $2.204$\\
        $c_{\perp,i}$ & $0.97$ & $1.879$\\
        $n_{i,\mathrm{in}}$ & $3.196$ & $1.63$\\
        $n_{i,\mathrm{out}}$ & $16.731$ & $-27.291$ \\\hline 
    \end{tabular}
    \tablefoot{Parameter definitions are found in the main text.}
\end{table}

\subsubsection{Galactic disk -- axisymmetric components}
\label{sec:disk}
The disk potential of our model takes the form of two exponential disk components with a hole in the centre, introduced to make room for the bar. We adopt a modified version of Equation 3 of \citet{McMillan2017} using an exponential vertical profile. We obtain
\begin{align}\label{eq:disk}
    \rho_\mathrm{disk} (R,z) &= \frac{\Sigma_1}{2h_1} \exp \Big(-\frac{R}{R_{\rm d,1}} -\frac{R_\mathrm{cut}}{R} - \frac{|z|}{h_1}\Big) \nonumber \\
    & + \frac{\Sigma_2}{2h_2} \exp \Big(-\frac{R}{R_{\rm d,2}}-\frac{R_\mathrm{cut}}{R} - \frac{|z|}{h_2}\Big)\;,
\end{align}
where $\Sigma_1 = 1.3719\times10^3$ M$_\odot$ pc$^{-2}$, $R_{d,1} = 2$ kpc, $z_1 = 300$ pc, $\Sigma_2 = 9.2391\times10^2$ M$_\odot$ pc$^{-2}$, $R_{d,2} = 2.8$ kpc, $z_2 = 900$ pc, and $R_\mathrm{cut} = 2.4$ kpc. The inner cutoff radius $R_\mathrm{cut}$, scale lengths $R_{\rm d}$ and surface density normalizations $\Sigma$ are obtained by fitting our model to the circular-velocity curves of \citet{Eilers2019} and \citet{Mroz2019} as shown in Figure~\ref{fig:rot_curve}, whereas the scale heights $h$ are fixed to the values from \citet{McMillan2017}, which are, in turn, obtained from SDSS star counts by \citet{Juric2008}. The parameterisation of the disk keeps the scale height fixed across all Galactic radii for simplicity, despite observations indicating the scale height decreases towards the Galactic centre. For example, $h(R=4\mathrm{ kpc}) \sim 180$pc for the thick stellar disk \citep{Wegg2015}.

In order to better represent the vertical acceleration towards the midplane of the Galaxy (for $z<400$ pc), we also include two gas disks in the potential, which we take from \citet{McMillan2017} without any further adjustments:
\begin{align}\label{eq:gas_disk}
    \rho_\mathrm{gas} (R,z) & = \frac{\Sigma_1}{4z_1}\exp\bigg(-\frac{R_{m,1}}{R} - \frac{R}{R_{d,1}}\bigg) \mathrm{sech}^2(z/2z_1)\nonumber \\
    &+ \frac{\Sigma_2}{4z_2}\exp\bigg(-\frac{R_{m,2}}{R} - \frac{R}{R_{d,2}}\bigg) \mathrm{sech}^2(z/2z_2)\;,
\end{align}
 where $\Sigma_1 = 53.1$ M$_\odot$ pc$^{-2}$, $R_{d,1} = 7$ kpc, $z_1 = 85$ pc, and $R_{m,1} = 4$ kpc represents the thick H\,\textsc{i} disk, whilst $\Sigma_2 = 2.18\times10^3$ M$_\odot$ pc$^{-2}$, $R_{d,2} = 1.5$ kpc, $z_2 = 45$ pc, and $R_{m,2} = 12$ kpc represents the thinner H$_2$ disk. Note that the gas disk in the hydrodynamical simulations in this paper is not self-gravitating and does not contribute to the potential; instead, these two gas disks are included as static components of the potential. It should also be noted that the gas disk potential does not contain a spiral perturbation, as we are interested in how the stellar potential affects the gas distribution in the simulations.

\subsubsection{Galactic disk -- spiral arms}
\label{sec:spiral-arms}
To generate the spiral arms of the Galaxy, we introduce a perturbation to the stellar disk in the following manner:
\begin{equation}\label{eq:spiral_arm}
    \rho_\mathrm{spiral} (R,z,\phi) = \rho_\mathrm{disk}(R,z) \cdot  \alpha\frac{R^2}{R_0^2} S(R,\phi).
\end{equation}
Here the perturbation strength increases quadratically with radius in order for the spiral arms to be strong enough in the outer regions of the Galaxy. \respB{For our fiducial model, we adopt an amplitude of $\alpha=0.36$ leading  to a spiral arm strength of $\sim 17$\% of the disk density at the solar radius of $R_0=8.179$ kpc \citep{GRAVITYCollaboration2019}.\footnote{We adopt the value suggested by the \citet{GRAVITYCollaboration2019} for comparison with other work in the literature, but we note that there is still some discussion about the exact value with a more recent estimate being $R_0=8.277 \pm 0.028$ kpc \cite[see Appendix C in][]{GRAVITY2021}.} This is somewhat larger than the 10\% value suggested by \citet{Eilers2020}, however, we note that we measure the dynamic response in the gas rather than the field star population, which is expected to be stronger due to the more dissipative nature of the ISM. We also note that we expect the impact of the choice of $\alpha$ to sensitively depend on the actual implementation of the thermal physics and chemistry in the code (see also Appendix~\ref{app:spiral_strength}). As outlined in Section \ref{sec:methods}, we here adopt a simple isothermal equation of state for the galactic gas. Consequently, we suggest to explore different values of $\alpha$  when accounting for the true multi-phase nature of the Galactic ISM and its multitude of heating and cooling processes (see e.g.\ \citealt{Tielens2005} or \citealt{Draine2011}).}

\begin{figure}
    \centering
    \includegraphics{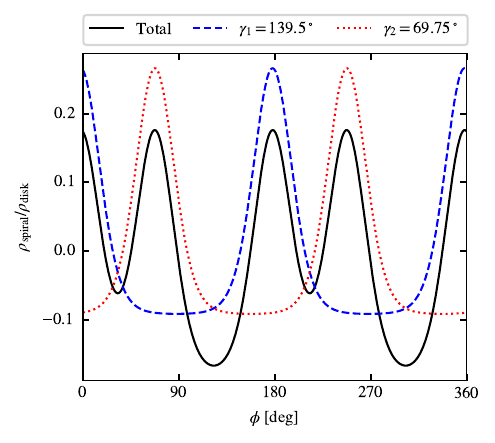}
    \caption{The strength of the spiral arms at solar circle, $R=8.179$ kpc, as a function of azimuth. Shown is the strength of the spiral arms in black, as well as the $f_{m_1,\gamma_1}$ and $f_{m_2,\gamma_2}$ components in blue and red respectively.}
    \label{fig:spiral_phi}
\end{figure}

For the shaping function, $S$, we make use of a logarithmic spiral arm potential with the width of the arm having a Gaussian profile. We take a modified form of Equation 8 of \citet{Junqueira2013}:
\begin{equation}\label{eq:shape}
\begin{split}
    S(R,\phi) = \sum_{k=1}^2 \bigg\{
      \exp\left( -\frac{R^2}{\sigma_\mathrm{sp}^2} \big[1-f_{m_k,\gamma_k}(R,\phi)\big] \right) \\
    - \exp{\left(-\frac{R^2}{\sigma_\mathrm{sp}^2}\right)}I_0\left(-\frac{R^2}{\sigma_\mathrm{sp}^2}\right) \bigg\},
\end{split}
\end{equation}
where
\begin{equation}\label{eq:log_sp}
    f_{m,\gamma}(R,\phi) = \cos \left(m(\phi+\gamma)-  \dfrac{m}{\tan(i)}\ln\left(\dfrac{R}{R_a}\right) \right)\;,
\end{equation}
and $i=12.5^\circ$ and $R_a = 9.64$ kpc, $m_1=m_2=2$, $\gamma_1 = 139.5^\circ$ and $\gamma_2=69.75^\circ$. $\sigma_\mathrm{sp} = 5$~kpc is the width parameter of the spiral arm which corresponds to a physical width of 1.082~kpc perpendicular to the spiral arm. The second term in Equation~\eqref{eq:shape} is used to normalise the spiral arm potential such that the monopole component is zero. Here, $I_0$ is the modified Bessel function of the first kind and of zeroth order. This resulting potential is a superposition of two pairs of $m=2$ spiral arms with equal amplitude \citep{Li2022}. We note that the spiral arm potential does not have an $m=4$ pattern due to the unequal angular separation between spiral arms. We opt for this kind of pattern as it allows for the angular separation to be adjusted as needed as the spiral arms of the Milky Way are not fixed to $90^\circ$ in angular separation \citep{Reid2019}. The shape and intensity of the spiral arms at $R=8.179$ kpc is shown in Figure \ref{fig:spiral_phi}.

\subsubsection{Dark matter halo}
\label{sec:DM-halo}
\begin{figure}
    \centering
    \includegraphics{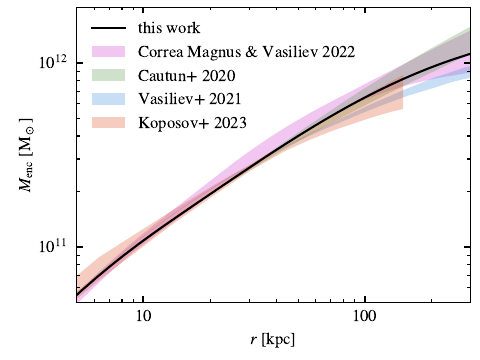}
    \caption{Enclosed mass profile of our fiducial potential (black), compared to the constraints from dynamical modelling of satellite galaxies (magenta: \citealt{CorreaMagnus2022}, green: \citealt{Cautun2020}) and streams (blue: \citealt{Vasiliev2021}, red: \citealt{Koposov2023}).}
    \label{fig:enc_mass}
\end{figure}

The dark matter halo component  follows a spherical \citet{Einasto1969} profile:
\begin{equation}\label{eq:dm}
    \rho_\mathrm{dm}(r) = \rho_0\exp\bigg[-\bigg(\frac{r}{a}\bigg)^{1/n}\bigg].
\end{equation}
The density normalisation, $\rho_0$, is determined by using the total mass of an Einasto potential:
\begin{equation}\label{eq:einasto}
    M = 4\pi\,\rho_0\,a^3\,n\,\Gamma(3n)\;,
\end{equation}
where the total mass is $M = 1.1\times10^{12}$ M$_\odot$, the Einasto index is $n=4.5$ and $\Gamma$ is the gamma function. The scale radius $a$ is related to the half mass radius $r_s$ by $a \approx r_s\,(3n-1/3)^{-n}$. In this case, the half mass radius is $r_s=96\,$kpc giving a scale radius of $a = 0.88$ pc. 
These parameters are optimised to simultaneously fit the circular-velocity in the inner region of the Galaxy and its mass distribution at larger distances determined from dynamical modelling of satellite galaxies and stellar streams \citep{Cautun2020,CorreaMagnus2022,Vasiliev2021,Koposov2023} (see Figure~\ref{fig:enc_mass}).

\subsection{Comparison with observations}
The parameters of the fiducial potential were optimised to satisfy a variety of recent observational constraints, as described below.

\subsubsection{Axisymmetric components}
We begin with the Galactic circular-velocity curve, as illustrated in Figure~\ref{fig:rot_curve} for different radial bins. 
The black line represents our total circular-velocity curve from the axisymmetrised potential, $v_c = \left(R\,{\partial\Phi_0}/{\partial R}\right)^{1/2}$, with the colored lines indicating the contributions from individual components based on the choice of parameters outlined above. Here the axisymmetrised potential is obtained from the monopole, $m=0$, component of the potential in which has been approximated by a Fourier or multipole expansion (see Section~\ref{sec:potential_method}). The resulting circular-velocity curve does not contain perturbations from the spiral arms nor contains the higher order terms needed to describe the full potential of the bar.

We make use of recent measurements of circular velocity data from \citet{Eilers2019}, obtained from red giant star observed with \textit{APOGEE, WISE} and \textit{Gaia}, and from \citet{Mroz2019}, obtained from Cepheid variable stars with \textit{Gaia}. Both are in a reasonable agreement with each other and provide a coverage of Galactocentric radius of $4 \lesssim R \lesssim 25$ kpc. For coverage within the solar circle, $R < R_0$, we make of use of the terminal velocities measurements from H\,\textsc{i} and CO observations \citep{Clemens1985,Fich1989,Burton1993,McClure-Griffiths2007}. We compare the peaks of the resulting longitude-velocity ($lv$) diagrams from our hydrodynamical simulations with the corresponding terminal velocity measurements of the Milky Way, as discussed in detail in Section~\ref{sec:results}. Here we opt to compare terminal velocities in the $lv$ diagram instead of the circular-velocity/rotation curves within the inner most $R<4$ kpc, as rotation curves obtained from observations in this region will include deviations due to the non-axisymmetric nature of the bar \citep{Chemin2015}. These deviations are not present in our axisymmetrised circular-velocity curve.

\begin{figure}
    \centering
    \includegraphics{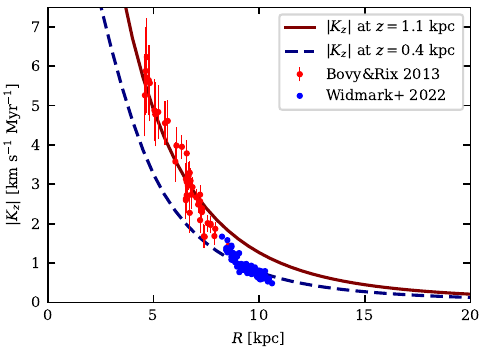}
    \caption{Vertical acceleration at $|z|=1.1$~kpc (solid red) and 0.4~kpc (dashed blue line). The observational constraints from \cite{Bovy2013} in the inner Galaxy at $|z|=1.1$~kpc and from \cite{Widmark2022a} in the outer Galaxy at $|z|=0.4$~kpc are plotted by red and blue points.}
    \label{fig:rot_surf}
\end{figure}

We also ensure that our potential satisfies observational constraints on the vertical acceleration at $|z|=1.1$~kpc derived in \citet{Bovy2013} from SEGUE kinematics of G-type dwarf stars, as well as the vertical acceleration at $|z|=0.4$~kpc obtained in \citet{Widmark2022a} from modelling the phase spiral in \textit{Gaia} EDR3 (see Figure~\ref{fig:rot_surf}). 

\subsubsection{Non-axisymmetric components}
\begin{figure}
    \centering
    \includegraphics{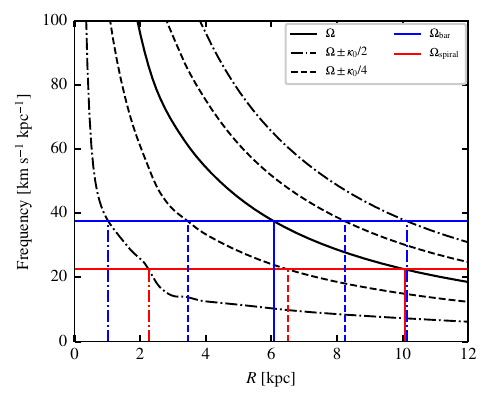}
    \caption{Frequency curve as a function of Galactic radius. The solid black line is the rotational frequency curve of the potential, whereas the dashed and dot-dashed lines are rotional frequency plus or minus 0.25 and 0.5 times the epicyclic frequency, $\kappa_0(R) = \sqrt{(2\Omega/R)\mathrm{d}(R^2 \Omega)/\mathrm{d} R}$. The horizontal blue and red lines are the pattern speed of the bar and  the spiral arms, respectively. The vertical dashed and dot-dashed lines are the corresponding resonances for  bar and spiral arms.}
    \label{fig:epicyc}
\end{figure}

Our potential has two rotating non-axisymmetric components, the bar and the spiral arms. We chose the pattern speeds of the bar and the spiral arms as $\Omega_\mathrm{bar}=-37.5$~km~s$^{-1}$~kpc$^{-1}$ \citep[e.g.][]{Sormani2015c,Sanders2019,Li2022,Clarke2022} and $\Omega_\mathrm{spiral}=-22.5$~km~s$^{-1}$~kpc$^{-1}$ \citep{Li2022}, respectively. We checked the consistency of these values by running a small parameter study with our potential to generate longitude-velocity $lv$ diagrams and comparing them to the spiral arm tracks presented in \citet{McClure-Griffiths2004}, \citet{Reid2016} and \citet{Reid2019}. For the sake of simplicity, we consider both non-axisymmetric components to experience solid body rotation. See Appendix~\ref{app:param} for the full details of the parameter study. The resonances for the potential can be found in Table~\ref{tab:resonance} and are illustrated in the frequency curves of Figure~\ref{fig:epicyc}. We find that for the pattern speeds we use, the outer Lindblad resonance of the bar coincides with corotation of the spiral arms at $\sim10.1$ kpc. The outer 4:1 resonance of the bar lies close to solar circle at $\sim8.2$ kpc. 

\begin{table}
    \caption{Location of resonances of the non-axisymmetric components of the potential.}
    \label{tab:resonance}
    \begin{tabular}{lccccc}
         Resonance & ILR & In~4:1 & CR & Out~4:1 & OLR  \\
         & (kpc) & (kpc) & (kpc) & (kpc) & (kpc)\\\hline
         Bar & 1.02 & 3.45 & 6.08 & 8.24 & 10.14\\
         Spiral arms & 2.36 & 6.50 & 10.07 & 13.06 & 15.89\\\hline\\[-0.1cm]
    \end{tabular}
    \tablefoot{
    Notation of resonances: ILR = Inner Lindblad, In~4:1 = inner 4:1, CR = corrotation, Out~4:1 = outer 4:1, OLR = outer Lindblad}
\end{table}

\section{Numerical simulations}
\label{sec:methods}
Here we briefly describe the numerical methods used to simulate the dynamical evolution of the ISM in our Milky Way analog. 

\subsection{Numerical hydrodynamics}
\label{sec:hydro} 

We solve the equations of hydrodynamics with the   moving-mesh code \textsc{Arepo} \citep{Springel2010}. For isothermal gas in two dimensions these are 
\begin{align}
&\frac{\partial \Sigma}{\partial t} + \nabla \cdot (\Sigma \mathbf{v}) = 0\;,\\
&\frac{\partial \Sigma \mathbf{v}}{\partial t} + \nabla \cdot (\Sigma \mathbf{v}\otimes \mathbf{v}) = -\nabla P -\Sigma\nabla\Phi\;,
\end{align}
where $\Sigma$, $\mathbf{v}$, and $P$ are  gas surface density, velocity and pressure, respectively. The simulations are two-dimensional. The pressure is related to the density via the equation of state,  $P = c_\mathrm{s}^2 \Sigma$, with the sound speed adopted as $c_\mathrm{s} = 10\,$km$\,$s$^{-1}$. The external potential $\Phi$ is given by our model for the Galactic potential, as explained in the next section; for ease of interpretation, we do not include the gas self-gravity, star formation or stellar feedback in our models. Note that by choosing a relatively large value for $c_{\rm s}$, we are implicitly accounting for some of the turbulent support of the gas disk, something that in reality would be provided by stellar feedback \citep{MacLow2004, Krumholz2005, McKee2007, KlessenGlover2016}. This assumption is justified in this case as our focus is on how the large scale dynamics impact kinematic distance estimates and not on the impact by turbulent motions.

\begin{figure}
    \centering
    \includegraphics{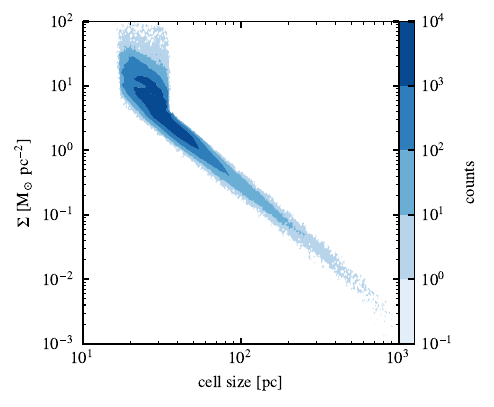}
    \caption{The relation between cellsize and density within our simulation. The cellsize in this case is the size of a square with the same area as the cell.}
    \label{fig:resolution}
\end{figure}

\textsc{Arepo} constructs a Voronoi tesselation, in which the mesh generating points are able to flow with the gas in the simulation, resulting in a quasi-Lagrangian approach to modeling the flow properties. We make use of an exact Riemann solver for isothermal flows, and the mesh can refine and derefine with the addition and removal of mesh generating points. This occurs when the mass of a given cell is a factor $\sim$2 larger or smaller than the cell target mass for the simulation ($M_{\rm target} = 2500$~M$_\odot$). The cell will either split or merge with another with addition or removal of a mesh generating point. For further details of the code base we use, see e.g.\ \citet{Tress2020a, Tress2020b}. We achieve a minimum value of $\sim40$ pc. Figure~\ref{fig:resolution} illustrates how our cellsize varies with density. The vertical spread seen at low cellsizes is a result of the minimum surface area of the cell being reached, which is set to 6 pc$^2$. This creates a limit on how small our cells can become by not allowing cells smaller than two times the minimum surface area to refine further.

\subsection{External Galactic potential}
\label{sec:potential_method}

\respA{\textsc{Arepo} allows one to include an external gravitational potential in the simulation, but the Galactic potential described in Section~\ref{sec:potential} is significantly more complex than the few built-in analytic models. We calculate the potential using the \textsc{Agama} library for stellar dynamics \citep{Vasiliev2019}, which, among other features, contains a powerful framework for constructing and evaluating arbitrarily complex potentials (including time-dependent features such as a rotating bar with a varying amplitude and/or pattern speed).
We created an interface between the two codes that make it possible to use any potential implemented in \textsc{Agama} as an external potential in \textsc{Arepo} (in addition to self-gravity of the simulated system, if the latter is turned on). Moreover, a very similar interface is provided for the \textsc{Gadget-4} code \citep{Springel2021}, which shares a common ancestry with \textsc{Arepo}; both interfaces, as well as the script for generating the Galactic potential from this study, are available in the latest version of \textsc{Agama}.}

\respA{The Galactic potential consists of two general-purpose expansions: \texttt{Multipole} for spheroidal density components, and cylindrical Fourier series (\texttt{CylSpline}) for disk-like components. Each of the two expansions is constructed from the sum of several density components, as detailed in Table~\ref{tab:components}; the mathematical details of these potential expansions can be found in the appendix of the \textsc{Agama} documentation \citep{Vasiliev2018}.}

\begin{table}
    \centering
    \caption{Expansion used for each component of the potential.}
    \begin{tabular}{lr}
        \hline
        Component & Expansion type  \\\hline
        Sgr~A*       & \texttt{Multipole} \\
        NSC          & \texttt{Multipole} \\
        NSD          & \texttt{Multipole} \\
        Bar          & \texttt{CylSpline} (1)  \\
        Stellar disk & \texttt{CylSpline} (1) \\
        Gas disk     & \texttt{CylSpline} (1) \\
        Spiral arms  & \texttt{CylSpline} (2)  \\
        DM halo      & \texttt{Multipole}\\\hline
    \end{tabular}
    \label{tab:components}
    \tablefoot{When present, spiral arms are represented by a separate \texttt{CylSpline} potential, since they rotate with a different pattern speed than the bar.}
\end{table}

\subsection{Initial conditions}\label{sec:ICs}
The initial conditions of our simulations are simple. We set up the surface density of the gas following an exponential profile similar to Equation~\eqref{eq:gas_disk}:
\begin{align}\label{eq:sim_disk}
    \Sigma(R) &
    =\Sigma_1\exp\bigg(-\frac{R_{m,1}}{R} - \frac{R}{R_{d,1}}\bigg) \nonumber \\ &+\Sigma_2\exp\bigg(-\frac{R_{m,2}}{R} - \frac{R}{R_{d,2}}\bigg)\;,
\end{align}
where the parameters for this gas disk are the same as those used for the gas disk potential (See Section~\ref{sec:disk}). We extend this gas disk to $R \sim 30$ kpc at which point we reduce the density significantly to prevent artifacts caused by periodic boundary conditions compromising the Galaxy itself.
For simplicity, we initialise the simulation with 250000 mesh generating points distributed uniformly across a (75 kpc)$^2$ box. The mesh is then relaxed with the \texttt{meshrelax} method within Arepo to reach our target mass of $2500\,$M$_{\odot}$ in which cells are refined or derefined according to the refinement/derefinement criterion mentioned previously. There is no hydrodynamics present in the meshrelax process and as such the gas is fixed until the process is complete.

The velocity of the gas is initialised to be the circular-velocity of the axisymmetric terms of the potential, $v_c = \left(R\,{\partial\Phi_0}/{\partial R}\right)^{1/2}$, which follows the same circular-velocity curve as Figure~\ref{fig:rot_curve}. The non-axisymmetric components of the bar and spiral arms are introduced linearly and gradually over the course of $150\,$Myr to avoid transients, as is customary in this type of simulations \citep[e.g.][]{Li2022}, making use of the time-dependent \texttt{Evolving} potential in \textsc{Agama}.

\section{Results}
\label{sec:results}
\subsection{The gas response}
In order to test kinematic estimates properly, the simulation box needs to be rotated such that the bar is in a similar position with respect to the Sun's position as it is for the Milky Way.
For each simulation output, we rotate the system so that the angle between the bar major axis and the Sun-Galactic centre line is 28$^\circ$ \citep{Bland-Hawthorn2016}. The Sun-Galactic centre distance is assumed to be $R_0=8.179\;$kpc \citep{GRAVITYCollaboration2019}. For each simulation we generate an $lv$ diagram assuming the Sun moves with a velocity equal to its circular-velocity in the $x$ direction, $v_x = v_c(R_0) = 229$~km~s$^{-1}$, and has no other velocity components. \respA{This value is obtained from the circular-velocity curve produced from the axisymmetric potential as detailed in Section~\ref{sec:potential}. It should be noted that the \citet{GRAVITYCollaboration2022} has given a more recent value of $R_0=8.277$ kpc, however we adopt the 2019 value for the calculations in this paper. Additionally, proper motion measurements of Sgr A$^*$ give $v_c(R_0)\sim250$ km s$^{-1}$ \citep[e.g.][]{Reid2020}, $\sim20$ km s$^{-1}$ greater than the value we have derived from our circular-velocity curve. This difference can be attributed to both the peculiar velocity of the Sun relative to the local standard of rest (LSR), and the spiral perturbations, which make the LSR velocity deviate from the axisymmetric circular speed. }

\begin{figure*}
    \centering
    \includegraphics{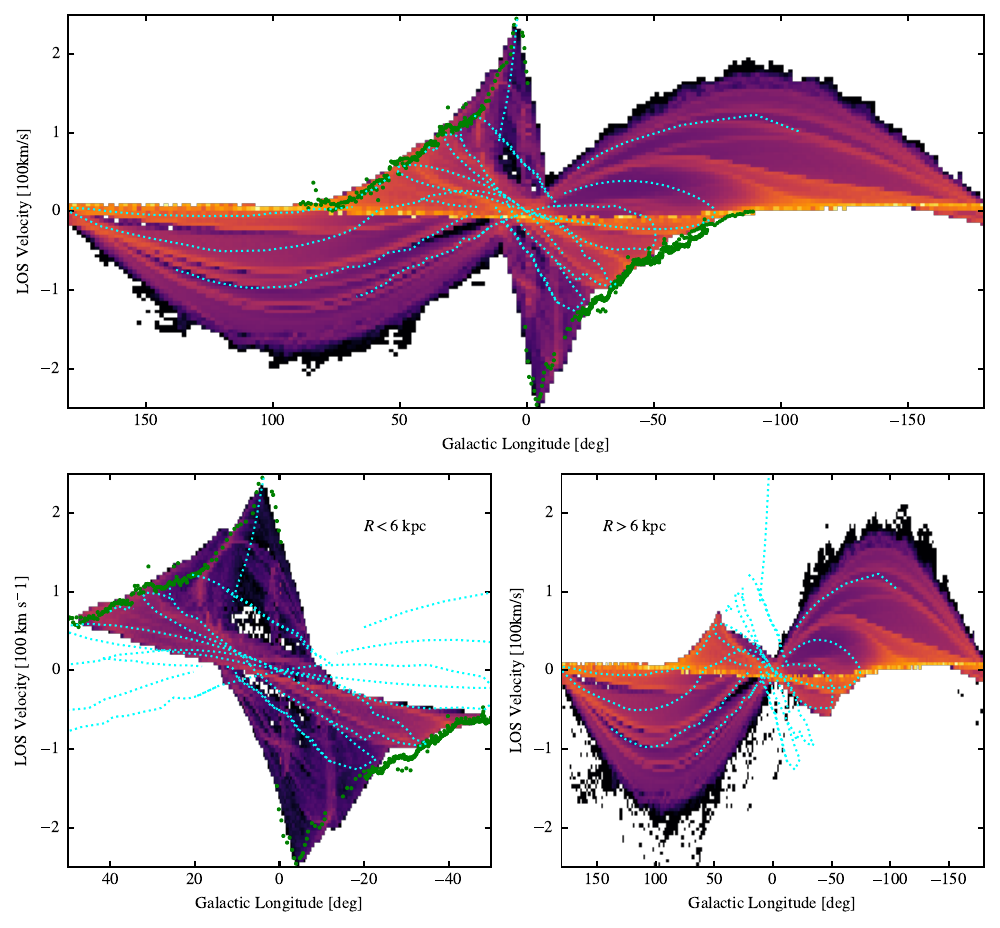}
    \caption{Longitude-velocity maps of the simulation observed from the Sun's position (placed at the origin of the overlayed coordinate system in Figure~\ref{fig:polar_dens}). The bar major axis is rotated by 28 degrees from the line of sight passing through the Galactic centre. Overlayed blue dashed lines are the spiral arm tracks of \citet{Reid2016,Reid2019} and \citet{McClure-Griffiths2004}. The green points are the terminal velocities from H\,\textsc{i} and CO observations. \textit{Top}: Full diagram. \textit{Bottom}: Zoom in between $l=-60^\circ$ \& $+60^\circ$ with the left plot only taking gas into account that lies at $R < 6\,$kpc from the centre and with the right one only considering gas at  $R > 6\,$kpc.}
    \label{fig:lv}
\end{figure*}

\begin{figure}
    \centering
    \includegraphics{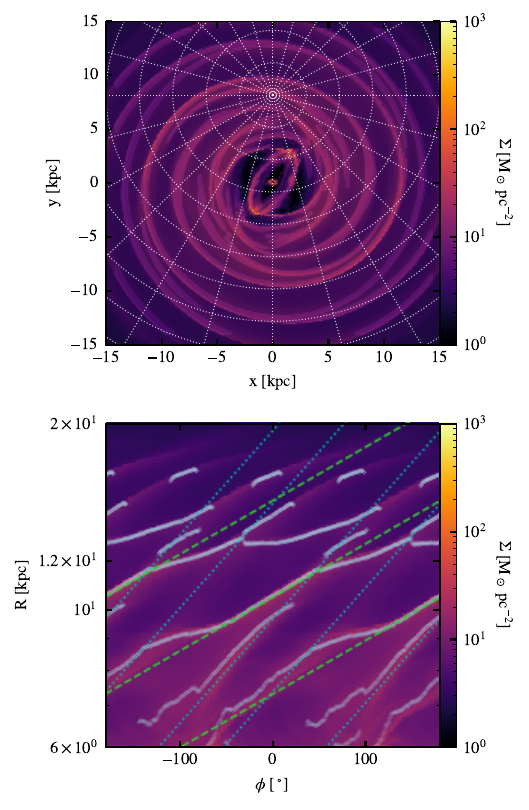}
    \caption{The top plot is the $x-y$ density projection of our simulation at $t=441$ Myr. \respA{Spiral arms are enhanced in this image using the mask generated by \textsc{filfinder} to brighten overdensities and darken other regions.} The bottom figure is the polar decomposition of the density map of the top figure focusing on the region with Galactocentric radius $6<R<20$ kpc. The blue dotted lines are the positions of the spiral arms according to the second half of Equation~\eqref{eq:log_sp}. The green dashed lines are the spiral arms generated by the rotation of the bar, with a pitch angle of $6.5^\circ$ around the outer Lindblad resonance, $R=10.14$ kpc. The light blue lines indicate the spiral arm pattern extracted with \textsc{filfinder}.}
    \label{fig:polar_dens}
\end{figure}

We compare the structures in the resulting $lv$ diagrams with the spiral arm tracks of \citet{Reid2016,Reid2019} and \citet{McClure-Griffiths2004}, as illustrated in Figure~\ref{fig:lv}. For the analysis presented in this Section, we select the system at 441~Myr. It is very similar to the observations in the $lv$ diagram, and the simulation at this point has been advanced for long enough that the non-axisymmetric components of the potential have had enough time to interact with the gas. The resulting density maps can be found in Figure~\ref{fig:polar_dens}.

From Figure~\ref{fig:lv} we find clear peaks in $lv$ diagram that are associated with the spiral arms generated from the underlying potential. The spiral arms generally trace the spiral arm tracks of \citet{McClure-Griffiths2004}, \citet{Reid2016} and \citet{Reid2019} in the regions outwith the Galactic centre ($|l|>50^\circ$). Towards the Galactic centre, comparing the spiral arms becomes difficult due to the perturbations generated by the bar. Some features are match, however there are features present in the $lv$ diagram that do not match with any track and vice versa.

Comparing the terminal velocities we find our simulation mostly fall within what is expected for the Milky Way, with the exception of two zones; one at $10^\circ< l < 30^\circ$ and the other at $-20^\circ<l<-10^\circ$. The LOS velocities are higher than that of the terminal velocities at these angles. This is a result of the circular-velocity curve at $sim3$ kpc being relatively steep in comparison to similar potentials \citep[See][]{Li2022}. This is a result of an overlap between the bar and stellar disk potentials, which causes a slight overdensity at this location. 

As expected, the potential of the bar strongly influences gas dynamics in the central region of the Galaxy. The gas here follows the typical $x_1$ orbits, a family of orbits elongated parallel to the major axis of the bar \citep{Contopoulos1989}, until it is shocked at the end of the bar, after which it flows inwards on nearly radial orbits. Eventually, the gas stabilises onto $x_2$ orbits, forming a ring of material at 220 pc from the centre of the Galaxy. This ring is the equivalent of the Central Molecular Zone (CMZ) and is consistent with the larger end of estimates for the Milky Way's CMZ \citep[e.g.][]{Henshaw2023}. 

Outside of the bar region ($R>5$~kpc), the gas forms a clear spiral pattern. It is rather complex and has two main components: a two-arm spiral caused by the rotation of the bar, and the four-arm structure created by the spiral component of the potential described in Section~\ref{sec:potential}. These two pattern rotate at different angular speeds, $\Omega_\mathrm{bar}=-37.5$~km~s$^{-1}$~kpc$^{-1}$ and $\Omega_\mathrm{spiral}=-22.5$~km~s$^{-1}$~kpc$^{-1}$ respectively, so they periodically interfere with each other. We plot the polar decomposition of the density map in the bottom plot of Figure~\ref{fig:polar_dens} to better illustrate the spiral patterns. Here, a straight line would be consistent with a logarithmic spiral. We observe two gradients of spiral structure: the underlying spiral arm structure from the potential (blue dotted), and an $m=2$ spiral being generated by the bar (green dashed) with an estimated pitch angle of $6.5^\circ$ near the outer Lindblad resonance of the bar. We note that the pitch angle value of the bar-generated pattern depends on the sound speed of the gas, as can be understood from the dispersion relation of spiral density waves in the tight-winding limit \citep[e.g.][]{Binney2008}. Whilst there are linear trends in the plot, there are deviations from the underlying spiral arm structure. These occur at the point where the two components intercept at $R\sim11$ kpc and at around the spiral arm crossing point, $R_{\rm cross}=9.64$ kpc, where the deviation is a bridging feature between the spiral arm and the bar induced spiral arm.

We extract the exact structure with the filament finding package \textsc{filfinder} \citep{Koch2015}. This package identifies structures from a 2D image using morphological techniques. Not only does the package provide the spines of the extracted structure, it also provides the masks of the extracted regions. We overlay the extracted spines in light blue in the bottom plot of Figure~\ref{fig:polar_dens} and we use the masks to  contrast the density of the simulation in the $x-y$ projection (Figure~\ref{fig:polar_dens}, top).

\subsection{Kinematic distance estimates}
The kinematic distance is calculated based on the assumption of purely circular orbits within the Milky Way. First, a rotation curve $v_c(R)$ is assumed. Then the Galactocentric radius of a given object is calculated by:
\begin{equation}\label{eq:gal_R}
    R = R_0\sin(l)\frac{v_c(R)}{v_0\sin(l)+v_\mathrm{los}}\;,
\end{equation}
where $l$ is the Galactic longitude of the object, $v_\mathrm{los}$ is its line-of-sight velocity, and $R_0$ and $v_0$ are the Galactic radius and circular-velocity of the Sun, respectively. 
For consistency, we use the circular-velocity curve generated by our potential as our rotation curve, as shown in Figure~\ref{fig:rot_curve}. 
Because $R$ appears on both sides of Equation~\eqref{eq:gal_R}, it is estimated through an iterative process.

The kinematic distance to the object is then obtained by
\begin{equation}\label{eq:kin_dist}
    d_k = R_0\cos l\pm\sqrt{R^2 - (R_0\sin l)^2}\;,
\end{equation}
It is possible for the kinematic distance estimate to return an undefined answer as a result of the argument inside the square root being less than zero. This occurs when
\begin{equation}
    v_\mathrm{los}^2 > v_\mathrm{term}^2 - 2v_0  (v_\mathrm{los}-v_\mathrm{term}) \sin l\;,
\end{equation}
where $v_\mathrm{term}$ is the terminal velocity along a given Galactic longitude and is given by: \citep{Burton1978}
\begin{equation}
    |v_\mathrm{term}| = v_c(R) - v_0 |\sin(l)|.
\end{equation}

In other words, the kinematic distance is indefinite when the observed $v_{\rm los}$ is not possible (e.g.\ too high) under the assumed rotation curve $v_c(R)$. In this case, the argument of the square root of Equation~\eqref{eq:kin_dist} is set to zero, which is equivalent to placing the object at the tangent point for a given $l$, where $R = R_0\sin(l)$. This is also equivalent to setting the velocity to the terminal velocity for a given $l$. 

Equation~\eqref{eq:kin_dist} can give two answers when observing inside the solar circle, resulting in the well-known kinematic distance ambiguity. 

\begin{figure*}
    \centering
    \includegraphics[width=0.98\textwidth]{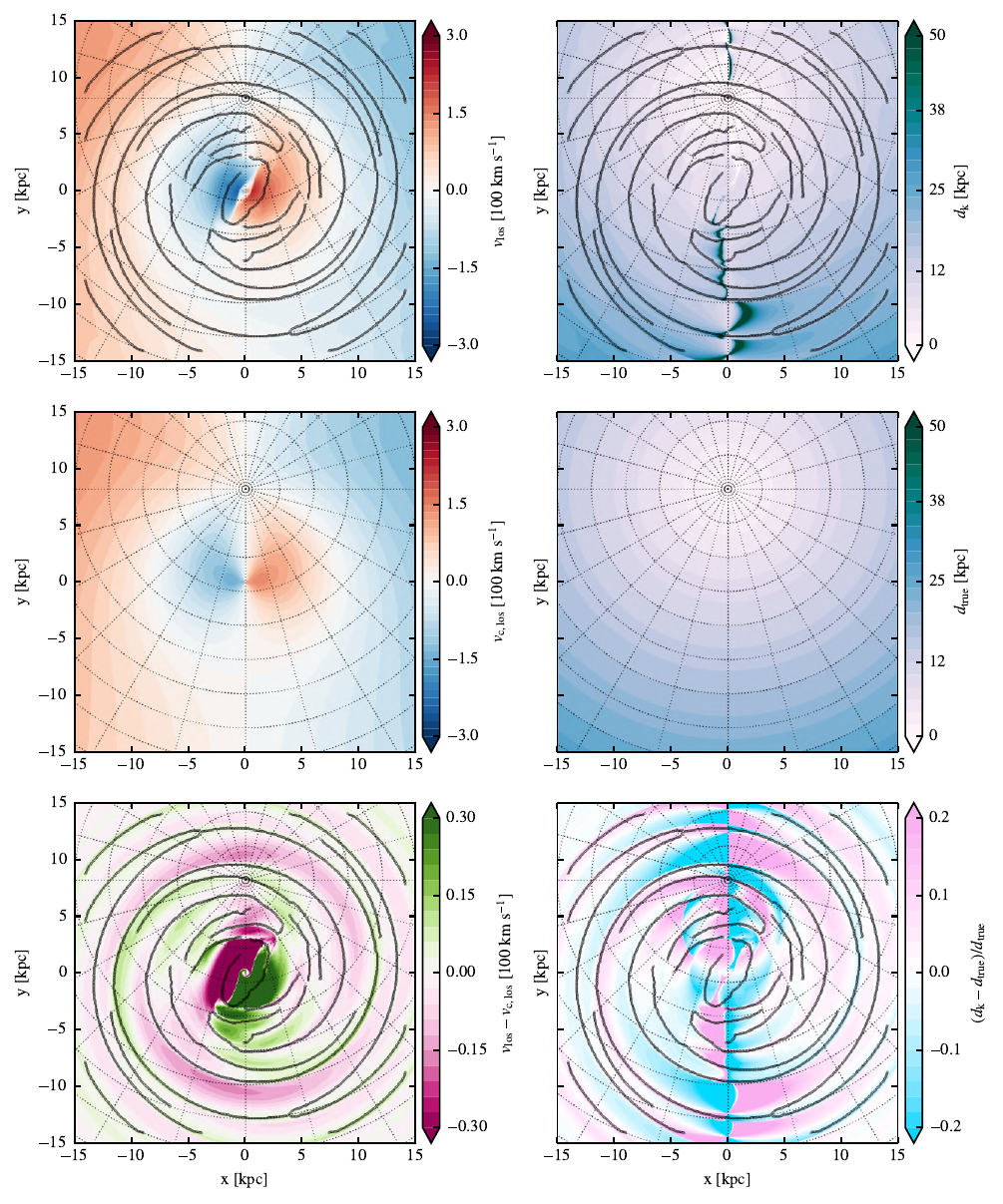}
    \caption{\respA{The left column presents the LOS velocity map of the simulation and the comparison to the equivalent if the gas followes circular orbits. The top plot illustrates LOS velocity map of the simulation. Similarly, the middle plot shows the map of the LOS velocity derived from  the base axisymmetric potential. The last plot shows the difference between simulation and axisymmetric LOS velocities; top plot minus the middle plot. The right column shows the} estimated and real distance maps in the simulation. The black lines indicate the spiral arm pattern extracted with \textsc{filfinder}. 
    The top plot shows the kinematic distance maps as estimated with Eq.~\eqref{eq:kin_dist}. The middle plot is the map of the true distances to the gas cells. The relative error between the kinematic and true distance is shown in the bottom plot.}
    \label{fig:d_compare}
\end{figure*}

For the sake of simplicity, we resolve the ambiguity by selecting the kinematic distance closest to the true distance value, \respB{but we note that this may be difficult to decide in real observations}.
\respB{In Figure~\ref{fig:d_compare} we present a face-on view of the disk and compare the data in our fiducial model with the ideal case of purely rotational motions. At the left we focus on the velocities and at the right we depict the corresponding distance estimates. In the top row of Figure~\ref{fig:d_compare}, we plot the line-of-sight velocity measured by an observer at the solar radius (left) and the KD estimates based on Equation~\eqref{eq:kin_dist}.} The kinematic distance map is not a smooth distribution with increasing radius from the Sun. \respB{We find deviations from the true values (as illustrated in the middle row Figure~\ref{fig:d_compare})} close to the perturbations caused by the spiral arms and the bar. We also observe quite large deviations close to $l=0^\circ$ and $l=180^\circ$. {\respA{This arises due to trigonometric effects: as $l$ tends towards $0^\circ$ or $180^\circ$, the value for the Galactic radius, $R$, obtained from Equation~\eqref{eq:gal_R} becomes ill-defined as both $\sin(l)$ and $v_\mathrm{los}$ tend to 0.}
\respB{These deviations are highlighted in the bottom row of Figure~\ref{fig:d_compare}}, where we map the difference in line-of-sight velicity and the relative error between the kinematic distance and the true distance.} What stands out is the large relative error in the solar neighbourhood. \respB{For objects within 500 pc of the Sun, the kinematic distance estimate is highly unreliable and exhibits a broad error distribution. Large systematic errors also occur close to the spiral arm perturbations and at the end of the bar, where the gas flows on $x_1$ orbits.}

When computing kinematic distances, observers typically avoid lines of sight within $\pm 15^\circ$ from the direction towards the Galactic centre and $\pm 20^\circ$ of the anti-centre due to high errors in these line of sights. \citep[e.g.][]{Anderson2012,Balser2015,Wenger2018}. {\respA{As $l$ tends towards $0^\circ$ or $180^\circ$, the value for the Galactic radius, $R$, from Equation~\eqref{eq:gal_R} becomes more difficult to estimate as both $\sin(l)$ and $v_\mathrm{los}$ tend to 0. On top of this, the Galactic bar also impacts estimates towards the centre due to the high level of asymmetry in the potential. }}

\begin{figure*}
    \centering
    \includegraphics{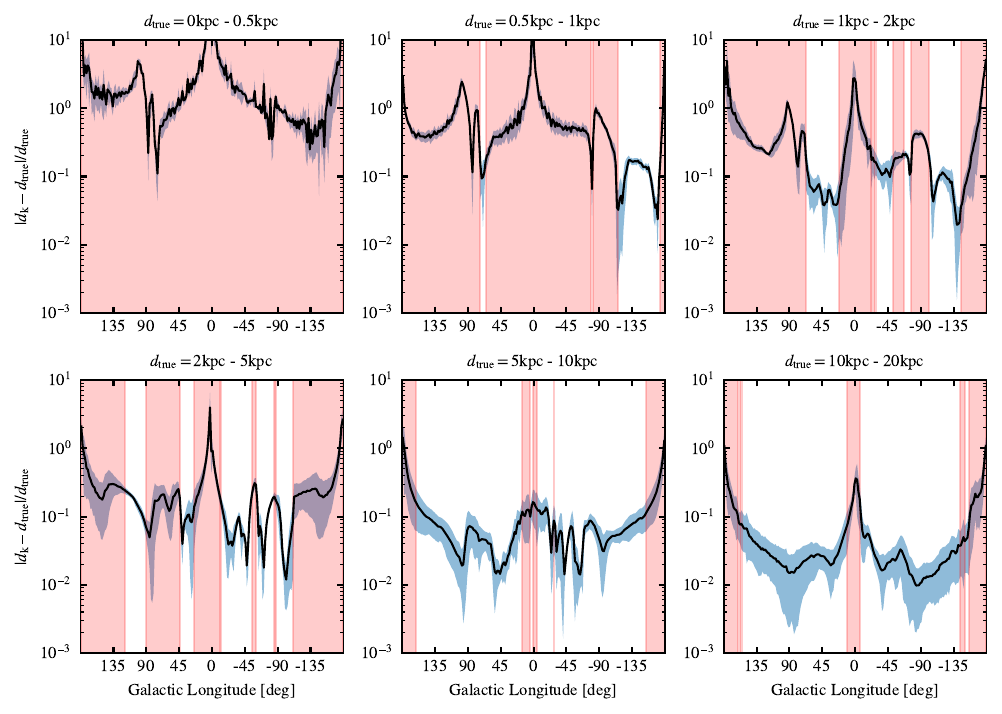}
    \caption{The black line represents the median absolute relative kinematic distance error along the line-of-sight of a given Galactic longitude. Each plot represents a sampling annulus centred on the Sun.  The blue shaded region is the median absolute deviation (MAD) of the error. The red shaded regions represent the Galactic longitudes where more than 20\% of the cells along the line-of-sight have a relative kinematic distance error of 27\% or more.
    }
    \label{fig:crown}
\end{figure*}

To give a more conclusive idea about which lines of sight to avoid, we compute the absolute kinematic distance error and plot the median and absolute median deviation (MAD) as a function of Galactic longitude in Figure~\ref{fig:crown}. The results are split into separate annuli around the Sun's position,
showing that the error in the kinematic distance remains high at distances closer to the Sun with some variation with Galactic longitude. We define a line of sight of avoidance for an annulus as the line of sight where more than 20\% of cells have an absolute kinematic distance error greater than 27\%. Our choice of this value is motivated by the study of \citet{Wenger2018}, who quantify the uncertainty in the kinematic distance inferred using the \citet{Brand1993} rotation curve (their method A) or the \citet{Reid2014} rotation curve (their method B) due to uncertainties in e.g.\ the solar Galactocentric radius and orbital velocity, the measured rotation curve, etc. The average absolute kinematic distance error they find when considering both models is 27\%. For lines of sight and locations where the systematic error for most points is less than this value, the kinematic distance method should be reliable. On the other hand, if a large fraction of points have systematic errors that exceed this value, this is a good indication that the kinematic distance method will not provide reliable results.
The lines of sight that should be avoided are shown in red in Figure~\ref{fig:crown}. For objects within 500$\,$pc of the Sun, 99.5\% of the full range of Galactic longitude should be avoided. This fraction remains above 56\% out to $5\,$kpc,  but drops to 14\% and 16\% for the $5-10\,$kpc annulus and the $10-20\,$kpc annulus, respectively. This suggests that the kinematic distance estimate is accurate for distances beyond $5\,$kpc from the Sun under our criterion for line of sight avoidance, 
but that for closer distances it should be used with great care.

\subsection{Location of kinematic distance errors}
So far we can see that velocity perturbations generated by the non-axisymmetric components of the potential can produce  highly inaccurate kinematic distance estimates along most lines of sight for objects close to the Sun. This now poses the question: where can one reliably use kinematic distances?

Back in Figure~\ref{fig:d_compare} we show the map of kinematic distance errors of our simulation with the density peaks extracted with \textsc{filfinder} overlayed onto the maps. The peaks lie close to the regions of low value for the kinematic distance error. However, this only applies to the spiral arm features, i.e.\ peaks outside of the bar region, from inspection. To further analyse this, we split the Galaxy into two regions: the bar region ($R<6\,$kpc) and the disk region ($R\geq6\,$kpc).

\subsubsection{Bar region}
\begin{figure}
    \centering
    \includegraphics{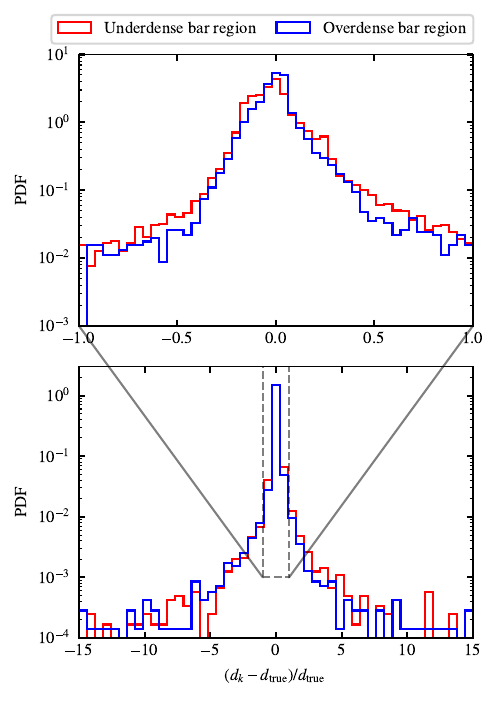}
    \caption{Probability distributions of the systematic kinematic distance errors within the overdensities in the bar region (blue) and the underdensities in the bar region (red).}
    \label{fig:bar_compare}
\end{figure}

For the bar region, we employ the mask generated by \textsc{filfinder} to identify the overdensities from our simulations and applying them to the  kinematic distance error map (Figure~\ref{fig:d_compare}), splitting the data into overdense regions and underdense regions. We plot the PDF of each region, respectively, and compare their distributions (Figure~\ref{fig:bar_compare}).

From visual inspection, we see that both distributions peak at around 0, with the underdense regions' distribution slightly wider than the overdense regions. Since the distribution is non-Gaussian, we look at the difference between quantiles to understand the width of the distributions. The difference between the upper and lower $20\%$ quantiles is 0.14 for overdense region, whilst it is wider with a value of 0.22 for the underdense region. Similarly, the values are 0.29 and 0.39 for the upper and lower 10\% quantiles, and 0.50 and 0.61 for the 5\% quantiles respectively. This suggests that outside of the overdense regions there is a higher probability of a large error and, by consequence, an increased probability of obtaining an incorrect distance via the kinematic distance method.
Going from our analysis of the error as function Galactic longitude (Figure~\ref{fig:crown}), this result is not too much of a surprise given that much of the bar's influence is lies within $l=\pm30^\circ$ which is typically a line-of-sight of avoidance for all distances away from the Sun. 

\subsubsection{Disk region}
As we have done previously, we split the kinematic distance maps of the disk region into spiral arm regions and interarm regions and plot the corresponding distributions of each. In this case we apply this to everything outside of $R=6\,$kpc.

\begin{figure}
    \centering
    \includegraphics{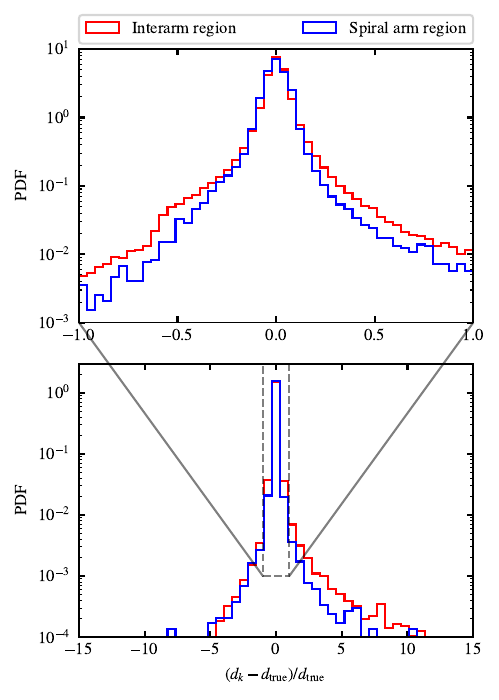}
    \caption{Probability distributions of the systematic kinematic distance errors within the spiral arm region (blue) and the interarm regions (red).}
    \label{fig:spiral_compare}
\end{figure}
From Figure~\ref{fig:spiral_compare} we find similarly shaped distributions in the errors for the Galactic disk as we did for the bar region. Once again we have the distribution peaking at 0 for both spiral and interarm regions, with a a wider distribution in the interarm regions.  Looking at the ranges between quantiles, the ranges are more similar between the spiral arm and interarm region. The difference between the upper and lower 20\% quantiles is 0.1 for both regions. For the difference in the 10\% quantiles the ranges are 0.17 and 0.2 for the spiral arm and interarm regions respectively. Similarly it is 0.26 and 0.39 for the 5\% qunatile difference. Although the difference in the quantiles are similar between the regions, the interarm regions has a wider distribution. This indicates that the interarm regions of the Galaxy have a higher probability of an incorrect kinematic distance, similar to that for the underdense regions of the bar though with a lower probability of a larger associated error. The narrower distribution of errors of the spiral arm region is a rather reassuring fact, because it implies that the kinematic distance method is more reliable in the regions of the Galaxy where most of the dense gas and subsequent star-forming regions are found. We look into the dynamics causing this result in the subsequent subsection.

It should be noted that in both overdense regions there is not much skewness in the distribution and, as such, there is no clear way to indicate whether if the kinematic distance would be under- or overestimated. We summarise the statistics generated from the systematic kinematic distance error distributions in Table~\ref{tab:dk_dist}.

\begin{table}[t]
    \centering
    \caption{\respA{The median and quantile differences of the distributions presented in Figures~\ref{fig:bar_compare} and \ref{fig:spiral_compare}.} }
    \begin{tabular}{l cccc}
       Region & Median &  \multicolumn{3}{c}{Quantile difference} \\\hline
         &  & 20\% & 10\% & 5\% \\\hline
        Overdense bar & 0.001 & 0.14 & 0.3 & 0.5\\
        Underdense bar & -0.018 & 0.22 & 0.39 & 0.61\\\hline
        Spiral arms & 0.001 & 0.1 & 0.17 & 0.26 \\
        Interarm & 0.004 & 0.1 & 0.2 & 0.39\\
    \end{tabular}
    \tablefoot{The quantile difference indicates the difference between upper and lower percent quantile, for example 20\% means the difference between the upper and lower 20\% quantile.}
    \label{tab:dk_dist}
\end{table}

\subsection{Relation to the velocity deviation}
\label{sub:vel-dev}
The key parameters that are needed for kinematic distance methods are the LOS velocity of the object and a rotation curve for the Milky Way. As mentioned previously, the Milky Way is not axisymmetric and as such there are deviations away from the rotation curve velocities. Quantifying the correlation between these deviations and the systematic kinematic distance errors can give an insight into how the velocity impacts the kinematic distance estimates.

\begin{figure}
    \centering
    \includegraphics{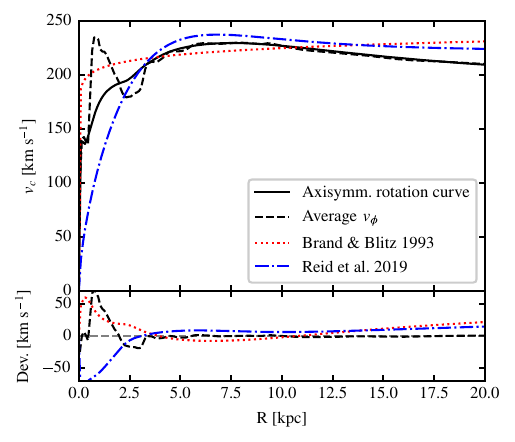}
    \caption{Comparison of rotation curve against the average azimuthal velocity (black solid and dashed line respectively). The red and blue lines are the rotation curve from \citet{Brand1993} and \citet{Reid2019}.}
    \label{fig:v_curve_compare}
\end{figure}

The deviations from rotation curve can be seen in the radial profile of the azimuthal velocity of the gas. We illustrate this in Figure~\ref{fig:v_curve_compare} where in the inner most 3 kpc we observed deviations up to 70 km s$^{-1}$. Between 3 kpc and 10 kpc there are small deviations from the rotation curve, on the order of few km s$^{-1}$, due to perturbations of the spiral arms in our system. We include the rotation curve from \citet{Brand1993} and the universal rotation curve {\respA{in physical units}} of \citet{Persic1996} with the updated parameters from \citet{Reid2019} as comparison to other rotation curves used in kinematic distance estimates.

\begin{figure}
    \centering
    \includegraphics{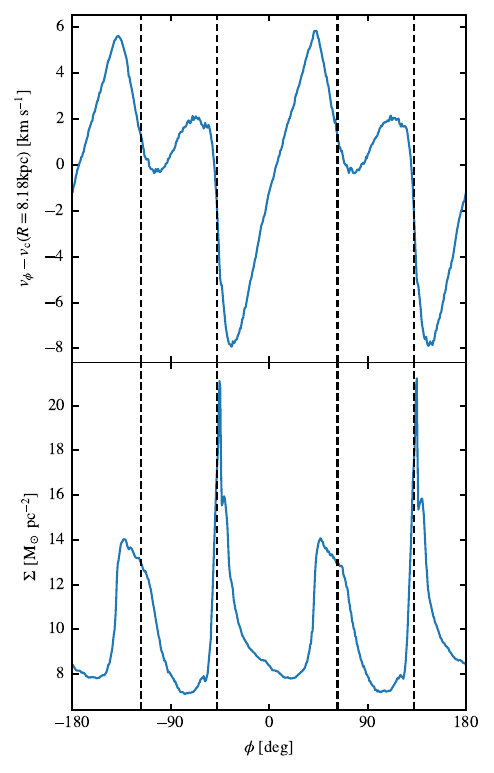}
    \caption{{\respA{The top figure illustrates the}} Galactic azimuthal profile of the difference between azimuthal {\respA{velocity, $v_\phi$ and circular velocity from circular-velocity curve, $v_c(R)$, near the solar circle at $R_0=8.179$ kpc. The bottom figure shows the azimuthal surface density profile}}. The vertical dashed lines are the potential minima of the underlying spiral arm potential. Rotation of the system is from \respA{left to right.}}
    \label{fig:vel_prof}
\end{figure}

We compute the line-of-sight velocity of our simulations and map it to the face down Milky Way view at the top {\respA{of the left column of Figure~\ref{fig:d_compare}}}. The non-axisymmetric perturbations of the potential are apparent here. Towards the bar region, we observe a sharp transition across the Galactic centre (as expected from the $lv$ diagram of Figure~\ref{fig:lv}), with perturbations of the spiral arms appearing as displacements in the contours in comparison to the {\respA{middle-left plot of Figure~\ref{fig:d_compare}}}; the LOS velocity maps for an axisymmetric potential.
When we look at what we expect from an axisymmetric potential, the transition towards the Galactic centre is not as strong and the contours of the line-of-sight velocity towards the outer Galaxy are smoother. When we subtract the two maps, we find that the largest deviations occur at the bar region with differences on the order of $100\,$km$\,$s$^{-1}$. 
Outside the bar region we find the line-of-sight velocity difference to be close to zero along the spiral arms of the Galaxy. 
Gas experiences an acceleration or deceleration as it flows into or out of the spiral arm, respectively as shown in Figure~\ref{fig:vel_prof}. This can cause shocks causing the gas to get denser as it leaves the potential minima, as in the case for two of our spiral arms. Additionally, the perturbations caused by the bar can also shock the gas and causes further perturbations. In our case, at around $R_0$, both the bar generated spiral and two of the spiral arms overlap creating large peaks in density.  However, unlike the other two spiral arms, the peak in density for these occur just before passing the potential minima of the spiral arm potential. \respB{Other more local sources of velocity perturbations come from expanding HII regions and supernova driven bubbles (see e.g.\ \citealt{Barnes2023} or \citealt{Watkins2023} for nearby galaxies, or \citealt{Zucker2022} for the solar neighborhood in the Milky Way).} 

All of this results in the largest deviations from the rotation curve to occur in the interarm regions which in turn causes shifts in the estimated kinematic distances, increasing their systematic error within these regions (see \respA{right-hand side of} Figure~\ref{fig:d_compare}). \respB{To be more quantitative, peculiar motions associated with the Galactic spiral arm structure of $\sim 6\,$km$\,$s$^{-1}$ in radial and $\sim 4\,$km$\,$s$^{-1}$ in tangential direction are reported by \citet{Reid2019}. We note that this is very much in line with the streaming motions of up to $6 - 8\,$km$\,$s$^{-1}$ illustrated in Figure~\ref{fig:vel_prof}.}

\begin{figure*}
    \sidecaption
    \includegraphics[width=12cm]{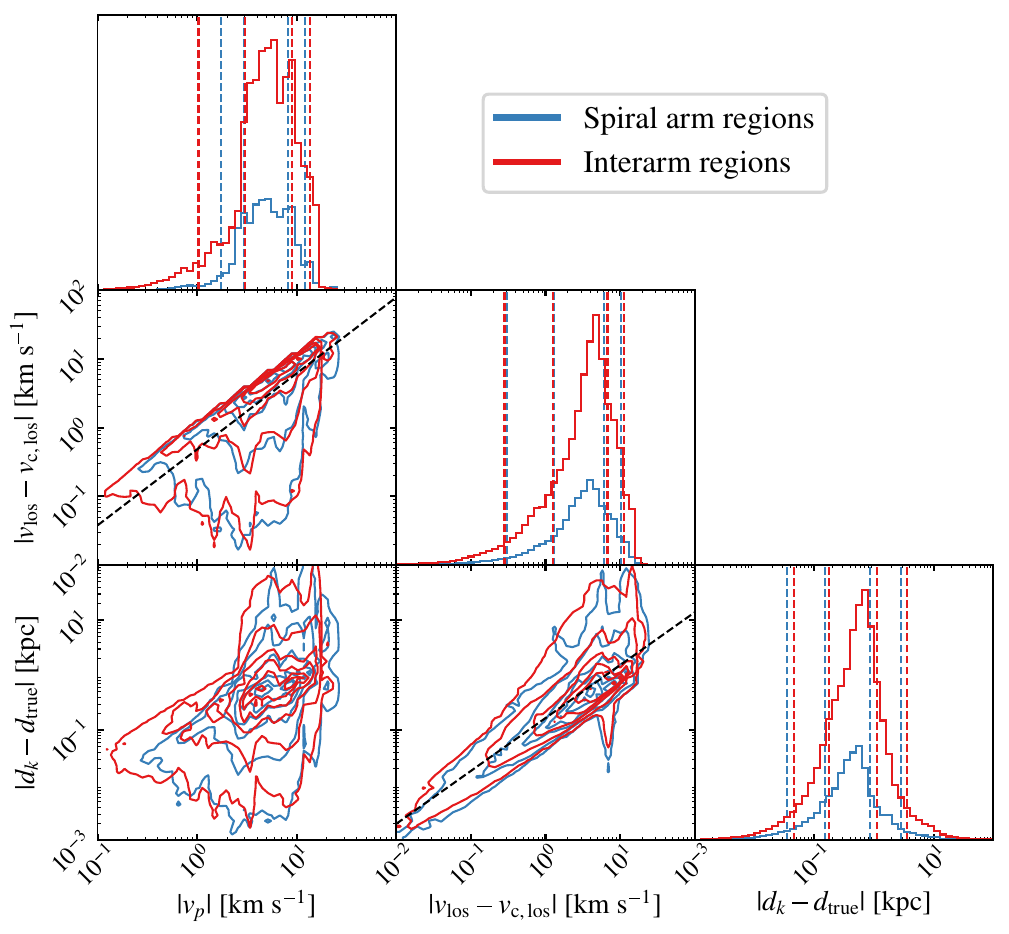}
    \caption{\respA{Distributions of the peculiar motions $|v_p|$, LOS velocity difference from circular motion $|v_\mathrm{los}-v_\mathrm{c,los}|$, and the difference between the kinematic and true distance $|d_k-d_\mathrm{true}|$. The dashed vertical lines on the histograms indicate the 5\%, 20\%, 80\% and 95\% percentiles of the distributions. There is a power-law correlation between the peculiar velocity and LOS velocity deviation of slope $\alpha=1.1$ (middle-left plot). Similarly, there is a power-law correlation between the LOS deviations from circular velocity and the resulting distance error (bottom-middle plot) with a slope  $\alpha = 0.94$.}}
    \label{fig:corner}
\end{figure*}

Given how large deviations of the line-of-sight velocity occur in the same regions where the systematic kinematic distance errors are highest, we look at the correlation between the \respA{Galactocentric radial velocity $v_r$} and the distance error. {\respA{Any deviation from circular motion is likely to occur due to motions in the radial direction.} In  Figure~\ref{fig:corner} \respA{we provide the distribution and correlations between absolute values of the peculiar motion $|v_p|$, the deviations from circular velocity $|v_\mathrm{los}-v_\mathrm{c,los}|$, and the difference between the true and kinematic distance $|d_k-d_\mathrm{true}|$, split between arm and interarm regions. There is a slight offset between the distributions for each region, with spiral arms having lower deviations and peculiar motions. As expected, we find a clear positive correlation between the local deviations from circular velocity and the resulting distance error. It can be described by a power-law with slope $\alpha=0.94$. However, we note that there are several orthogonal spurs in the 2D distribution that do not follow this trend. Additionally, we observe a power law correlation between the peculiar velocities and the difference in LOS velocity. Here, the slope is $\alpha=1.1$. Despite the positive correlations with the LOS velocity deviations, there is no clear correlation between the distance deviation and the peculiar motions.}}

\section{Discussion}
\label{sec:discussion}

\subsection{Implications}\label{sec:implications}
From our analysis, we find that kinematic distance estimates are most unreliable close to the Sun and along Galactic longitudes towards the Galactic (anti-)centre. 
Additionally, there is a  higher deviation within interarm regions than within spiral arms. This implies that clouds within underdense regions of the galaxy are more likely to have an incorrect distance estimate from the kinematic distance method.

An observational study carried out by \citet{Wenger2018} compared the distance estimates obtained by the parallax method to those obtained with the kinematic distance method. By treating the parallax distance as the true distance, they found   distance deviations of $\pm 40\%$ for their entire sample (see their Table 4). 
\respB{This is in line with the findings reported by \citet{Sofue2011} and our analysis here. As a cautionary note, we reiterate} {\respA{that our simulations do not contain any additional velocity perturbations, such as could be driven by stellar feedback (see also Section \ref{sec:caveats}), and so our numbers are likely to be lower bounds. However, to quantify this effect requires more detailed simulations that include all relevant physical processes self-consistently \cite[for a preliminary discussion see e.g.][]{tress2024}.}}

\begin{figure*}
    \centering
    \includegraphics[width=\textwidth]{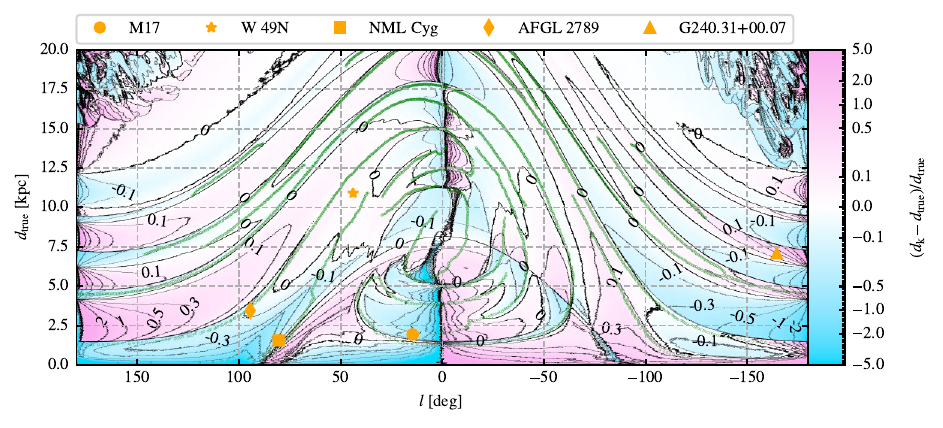}
    \caption{Map of the {\respA{relative}} kinematic distance error as a function of Galactic longitude and true distance. The contours are placed at $0, \pm0.1, \pm0.3, \pm0.5, \pm1, \pm1.5$ and $\pm2$. The green lines indicate the spiral arms extracted with \textsc{filfinder}. \respB{Overplotted in orange are a selection of known sources for which we compute the true distance $d_{\rm true}^{\,\sf w}$ from the parallax measurements of \citet{Wenger2018}}.}
    \label{fig:ld_err}
\end{figure*}

To help give an idea as to what implications this has for the observations, we generate a longitude-distance map of our kinematic distance errors in Figure~\ref{fig:ld_err}. \respB{We overplot some of the sources listed by \citet{Wenger2018} to indicate where real objects would lie on the map, taking the values derived from maser parallax measurements as true distances. We provide a quantitative comparison in Table\ \ref{tab:dist_compare} and find that in some instances the errors are similar, within a factor of two or less. However, in others they are quite different. For example, the error associated with AFGL 2789 has an error on the order of $-8\%$ in our system but is around $\sim50\%$ from observations.}
{\respA{We note that our hydrodynamic model for the response of the gas to the Galactic potential is chosen to be highly idealised, and for example include neither star formation nor stellar feedback (see the discussion in Section \ref{sec:caveats} below). The local gas flow in spiral arms is therefore approximate and we expect more stochastic behavior in more realistic simulations \cite[see e.g.][]{Tress2020a,Tress2021}. As AFGL 2789 is located within the Perseus spiral arm \citep{Oh2010} this could explain the deviation between our estimate and that of \citet{Wenger2018}. }}

\begin{table*}
    \centering
    \caption{\respA{Kinematic distances and errors of five well-studied sources in the Galactic disk.}}
    \begin{tabular}{lcccccc}
    Source & $l$ & $d_{\rm true}^{\,\sf w}$ & $d_{\rm k}^{\,\sf w}$ &$d_{\rm k}$ & $(d_{\rm k}^{\,\sf w}-d_{\rm true}^{\,\sf w})$ & $(d_{\rm k}-d_{\rm true}^{\,\sf w})$\\
    &(degrees)&(kpc)&(kpc)&(kpc)&$/ d_{\rm true}^{\,\sf w}$ & $/ d_{\rm true}^{\,\sf w}$\\\hline
         M17 & 15.03 & 1.97 & 2.33 & 1.10 & 0.18 & -0.44  \\
         W49N & 44.20 & 10.93 &  11.52 & 11.15 & 0.05 & 0.02\\
         NML Cyg & 80.80 & 1.6 & 1.33 & 1.32 & -0.17 & -0.18\\
         AFGL 2789 & 94.60 & 3.49 & 5.48 & 3.21 & 0.57 & -0.08\\
         G240.31+00.07 & -163.84 & 7.11 &  5.75 & 9.17  & -0.19 & 0.29\\\hline
    \end{tabular}
    \label{tab:dist_compare}
    
    \tablefoot{{\respB{For five Galactic sources (column 1) we list the longitude $\ell$ (column 2) and the true distance $d_{\rm true}^{\,\sf w}$ (column 3) derived from the maser parallax measurements reported by \citet{Wenger2018} as well as their KD estimate $d_{\rm k}^{\,\sf w}$ (column 4) for comparison with the value $d_{\rm k}$ from our model. We also provide the corresponding relative KD errors in columns 6 and 7.}} 
    Despite the simplicity of the physical processes included in our simulation, we {\respB{find  that our approach}} reproduces the observational data reasonable well.}

\end{table*}

\subsection{Caveats}
\label{sec:caveats}

There are a few limitations to bear in mind when considering the maps of kinematic distance errors derived from our simulations. First, as mentioned in Section~\ref{sec:methods}, the simulations performed here are 2D dimensional. This corresponds to the assumption that the gas in our simulation is integrated vertically, along the $z$-axis. The acceleration of the gas due to the potential is computed as if the gas lies in the midplane of the Galaxy ($z=0$).
This completely neglects the 3D structure of the Galaxy and vertical motions present within the gas. This additional component will impose changes to Equations~\eqref{eq:gal_R} and \eqref{eq:kin_dist} with the introduction of additional $\cos{(b)}$ terms.  With the perturbations induced by the potential, the gas can also experience changes in the $z$-component of the velocity as it travels in and out of a spiral arm. This will impact the LOS velocity of the gas and the resulting kinematic distance. However, quantifying the size of this effect is beyond the scope of this paper.

Second, the simulations presented here are idealised with only isothermal hydrodynamics and an external galactic potential. We note that the adopted value of $c_\mathrm{s} = 10\,$km$\,$s$^{-1}$ implies that we use an effective sound speed, which includes a strong turbulent component providing additional support of the gas disk \citep{MacLow2004, Krumholz2005, McKee2007, KlessenGlover2016}. Besides a self-consistent treatment of stellar feedback, more realistic models should include time-dependent chemistry \cite[e.g.][]{Glover2010}. Altogether, we would expect the sound speed to vary across the Galaxy, causing the structure of the Galaxy to alter. Indeed, parameter studies with different sound speeds have shown that spiral arms tend to get wider with increasing $c_\mathrm{s}$ \citep{Li2022}, whereas the size of the CMZ becomes smaller \citep{Sormani2015,Sormani2023}. 

The spiral arm potential we present in this paper is not an exact match to the spiral arm pattern of the Milky Way. It is well noted that the spiral arm shape of the Milky Way is not regular, with differences in phase angles between spiral arms and pitch angle changes along the spiral arm \citep[see][] {Reid2016,Reid2019}. These irregularities are hard to model when constructing the potential so approximations are required. Here, we keep the pitch angle of the spiral arms the same and try to account for the change in phase angle with the two pairs of Gaussian shaped spiral arms (see Section~\ref{sec:spiral-arms}). Additionally, perturbations generated by the interaction with orbiting satellite galaxies can impact the velocity space of the Galaxy, creating wave-like oscillations throughout the Galaxy \citep{Khanna2019}. These differences in structure will make some of the kinematic distance estimates in our simulation very different from what they would be for the Milky Way, as illustrated in Section~\ref{sec:implications}.

Turbulent motions induced by physics such as self-gravity, stellar and supernova feedback would contribute to the velocity dispersion of the system. The effects of self-gravity can add an additional $\sim2-5$ km s$^{-1}$ to the velocity dispersion for axisymmetric systems but can be as high as $\sim10$ km s$^{-1}$ for non-axisymmetric systems such as the one presented in this paper \citep{Wada2002}. Despite this, supernova feedback is believed to give the largest contributions to the velocity dispersion on large scales, potentially producing a velocity dispersion of as much as $\sim 10$ km s$^{-1}$ across hundreds of parsecs \citet{Lu2020}. This can result in the line-of-sight velocity deviating from the values derived here by a similar amount, causing kinematic distance estimates to deviate further from the true value. 

Kinematics distance are normally computed with one of two Galaxy rotation models, the rotation curve of \citet{Brand1993} and the universal rotation curve of \citet{Persic1996} with updated parameters from \cite{Reid2014}. Both of these rotation curves have been obtained from the gas within the Milky Way; the former making use of H\textsc{ii} regions and  H\textsc{i} tangent point data, whilst the latter makes use of maser parallaxes. We do see differences between the rotation curves presented in these papers and our values, since our potential is modelled based on the rotation curves from stellar data, as shown in Figure~\ref{fig:v_curve_compare} \citep{Mroz2019,Eilers2019}. There is a difference on the order of up to $\sim10$ km s$^{-1}$ between the rotation curve within the disk of the Galaxy, and larger deviations within the inner most 3 kpc. An investigation into how these differences in the standard rotation curves can impact kinematic distance is beyond the scope of this paper, but does warrant future investigation.

\respA{In addition, we note that the simple assumption of the Milky Way being close to dynamical equilibrium requires revisions. Besides the local velocity fluctuations due to stellar feedback, as discussed above, there is increasing evidence that our Galaxy is significantly perturbed on global scales. One of the main results of the \textit{Gaia} satellite has been to reveal that the Milky Way's disk is much farther from equilibrium than previously expected. It is subject to strong perturbations from satellite galaxies, such as the Magellanic Clouds and the Sagittarius dSph, that cause it to depart significantly from axisymmetry \cite[see e.g.][]{Antoja2018, Antoja2022,Widmark2022b}, and generate warps in the outer disk or even lead to a displacement of the baryonic component with respect to the global dark matter distribution \cite[e.g.][]{Laporte2018, Laporte2018b, Yaaqib2024, Chandra2024}. Future estimates of local distance estimates, especially for the outer disk, need to take this into consideration.}

\section{Conclusions} 
\label{sec:conclusion}
In this paper,  we have presented a realistic analytic potential for the Milky Way. It contains density profiles for all major mass component of the Galaxy. These are the supermassive black hole in the very centre, the nuclear stellar cluster and nuclear stellar disk, the Galactic bar, the Galactic disk, which we  split into  axisymmetric components for field stars and gas and a spiral arm component for the field stars only, and finally an extended dark matter halo that dominates the potential at large distances. These are introduced and fitted to the observational constraints, such as the rotation curve and the terminal velocities, in Section~\ref{sec:potential}. 

We also described how the new analytic potential is implemented in the moving-mesh code {\sc Arepo} using the \textsc{Agama} framework, as outlined in detail in Section \ref{sec:methods}. We made use of 2D hydrodynamic simulations to investigate how robustly the axisymmetric assumption holds for kinematic distance estimates. For this, we place an observer at $R_0=8.179$ kpc with the bar angled at 28$^\circ$, generate kinematic distance estimates to each of the gas cells present within our simulations and compute the systematic errors for each, $(d_k-d_\mathrm{true})/d_\mathrm{true}$. As discussed in Section \ref{sec:results} we found that the errors are high close to the Sun, with values reaching >50\% on average for any sources with 1 kpc. Along with proximity, we found that errors also reach these values when viewing towards the Galactic centre and anti-centre, $l=0^\circ$ and $l=180^\circ$, respectively. \respB{We report pecular motions associated with the Galactic spiral arm structure of up to $6 - 8\,$km$\,$s$^{-1}$, which is in line with the values inferred from maser observations in the Milky Way.}

When considering both Galactic longitude and distance, there are certain lines-of-sight that result in higher errors in addition to those previously mentioned for specific distance ranges. We identify these regions as zones of avoidance for the application of the kinematic distance method.  We also compare our results with the distance estimates of some well-studied molecular clouds and find in general good agreement (Section \ref{sec:discussion}).

In summary, the extraction of the velocity perturbations in our simulation has allowed us to determine what impact the potential has on systematic errors in the kinematic distance estimate. We find that within the spiral arms of the Galaxy, the kinematic distance errors are low as the gas lies within the local potential minima. Consequently, the line-of-sight velocity of the gas is close to what is expected for the axisymmetric version of our potential. We expect clouds within the spiral arms of the Milky Way to have low systematic kinematic distance errors. Conversely, the interarm regions present the largest deviation in both the kinematic distance and the line-of-sight velocity for a given Galactic radius. This is caused by the gas being sped up or slowed down as it travels into or out of the spiral arm. Additionally, we discovered a power law relation between the systematic kinematic distance error and the difference between the line-of-sight velocity and the projected circular-velocity. 

We conclude that the assumption of axisymmetry for the kinematic distance method can result in large systematic deviations depending on where a source is situated within the Milky Way. These deviations can alter derived values that depend on distance and, as such, the corresponding systematic errors should be accounted for in these derived values.

The interface between \textsc{Agama} and \textsc{Arepo}/\textsc{Gadget-4} codes, along with the scripts for constructing the potential from Section~\ref{sec:potential} and for running $N$-body simulations with these codes in the external potential, are included in the \textsc{Agama} repository\footnote{\url{https://github.com/GalacticDynamics-Oxford/Agama}}.

\begin{acknowledgements} 
\respB{We thank the referee for insightful commments that helped to improve the clarity of this paper. }
This investigation is funded by the European Research Council under ERC Synergy Grant ECOGAL (grant 855130). The authors gratefully acknowledge computing resources provided by the Ministry of Science, Research and the Arts (MWK) of the State of Baden-W\"{u}rttemberg through bwHPC and the German Science Foundation (DFG) through grants INST 35/1134-1 FUGG and 35/1597-1 FUGG. They also acknowledge data storage at SDS@hd funded through grants INST 35/1314-1 FUGG and INST 35/1503-1 FUGG. RSK furthermore acknowledges financial support from the Heidelberg Cluster of Excellence STRUCTURES in the framework of Germany's Excellence Strategy (EXC-2181/1 - 390900948), and from the German Ministry for Economic Affairs and Climate Action in project MAINN (funding ID 50OO2206).
The authors gratefully acknowledge the scientific support and HPC resources provided by the Erlangen National High Performance Computing Center (NHR@FAU) of the Friedrich-Alexander-Universität Erlangen-Nürnberg (FAU) under the NHR project a104bc. NHR funding is provided by federal and Bavarian state authorities. NHR@FAU hardware is partially funded by the German Research Foundation (DFG) – 440719683. \respA{MCS acknowledges financial support from the European Research Council under the ERC Starting Grant ``GalFlow'' (grant 101116226) and from the Royal Society (URF\textbackslash R1\textbackslash 221118). EV acknowledges support from an STFC Ernest Rutherford fellowship (ST/X004066/1).}
NB acknowledges support from the ANR BRIDGES grant (ANR-23-CE31-0005).
\respB{RSK also thanks the Harvard-Smithsonian Center for Astrophysics and the Radcliffe Institute for Advanced Studies for their hospitality during his sabbatical, and the 2024/25 Class of Radcliffe Fellows for highly interesting and stimulating discussions. JG is a member of the International Max Planck Research School for Astronomy and Cosmic Physics at the University of Heidelberg (IMPRS-HD).} 
\end{acknowledgements}


\bibliographystyle{aa}
\bibliography{ref}



\begin{appendix}
\section{Spiral arm strength}\label{app:spiral_strength}

We perform additional simulations to determine \respB{the impact of our choice of the spiral arm strength factor} $\alpha$ in Equation~\eqref{eq:spiral_arm}. \respB{Specifically, we investigate three different values, $\alpha=0.204$, $0.408$ and $0.612$, and find that they lead to a peak density contrast of 10\%, 20\% and 30\% at the solar circle at $R_0=8.179$ kpc, respectively. We caution that \citet{Eilers2020} suggest a density contrast of $\sim 10$\% for the vicinity of the Sun, and note that the fiducial case of $\alpha=0.36$ introduced in Section \ref{sec:spiral-arms} gives a density contrast of 17\%. However, we also point out that it leads to peculiar gas motions consistent with the maser data (as discussed in Section\ \ref{sub:vel-dev}, see also \citealt{Reid2019} or \citealt{Baba2009}). This is the main reason for our choice.} 

\respB{We also note that the dynamic response ISM to spiral potential perturbations strongly depends on the thermodynamic properties of the medium. Our fiducial value of $\alpha=0.36$ gives good results for our highly simplified description of the Galactic ISM, where we adopt a  isothermal equation of state, ignore self-gravity in the gas and consequently do neither treat star formation nor stellar feedback, and neglect the potential impact magnetic fields or cosmic ray physics  (for details of the numerical set-up, see Section\ \ref{sec:hydro}. This puts us in the position to study the gas response to the external potential without the need to correct for additional physical processes. We note that the inclusion of appropriate heating and cooling processes as well as star formation and stellar feedback is expected to increase the local deviations from circular velocity, and hence, the overall velocity dispersion. We speculate that simulations that remain close to the observed level of peculiar gas motions \citep[e.g.][]{Baba2009,Reid2019} are likely to require a smaller $\alpha$ and reach spiral arm density contrasts more in line with the values proposed by \citet{Eilers2020}. A detailed analysis of these effects will be subject to a follow-up investigation.}

\begin{figure}
    \centering
    \includegraphics{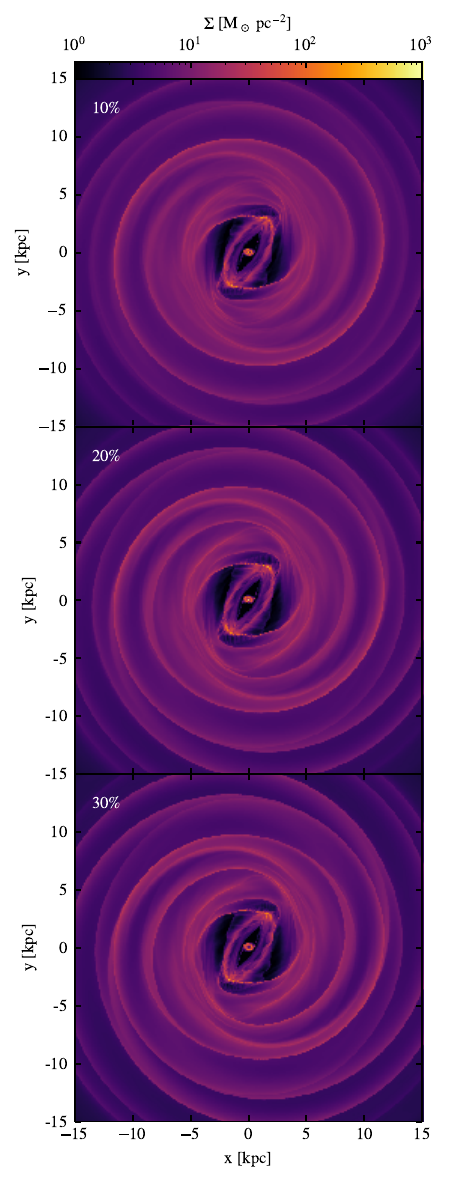}
    \caption{The density maps of simulations of differing spiral arm strength at $t=441$ Myr. The strength of the spiral arm pertibation increase from 10\% to 30\% stellar disk density at $R=8.179$ kpc from top to bottom. }
    \label{fig:spiral_strength}
\end{figure}

\section{Parameter study}
\label{app:param}
We perform a simple parameter study with the potential to understand how varying certain parameters impacts the $lv$ diagram. This involves multiple simulations with the changed parameters. The parameters we consider are number of spiral arms ($n$), pitch angle of the spiral arms ($i$), spiral arm pattern speed ($\Omega_\mathrm{spiral}$) and bar pattern speed ($\Omega_\mathrm{bar})$. The values considered are indicated in Table~\ref{tab:param_study}. We consider two sets of parameter studies, one for each spiral arm number, in which we vary one of the other parameters and keep all other fixed to the fiducial values. Our fiducial parameters are based on the same parameters in \citet{Li2022}. This allows us to investigate how each individual parameter alters the $lv$ diagram and later to vary the parameters to obtain an appropriate approximate match between the observed spiral arms and those in our simulations in $lv$ space.

\begin{figure*}
    \centering
    \includegraphics{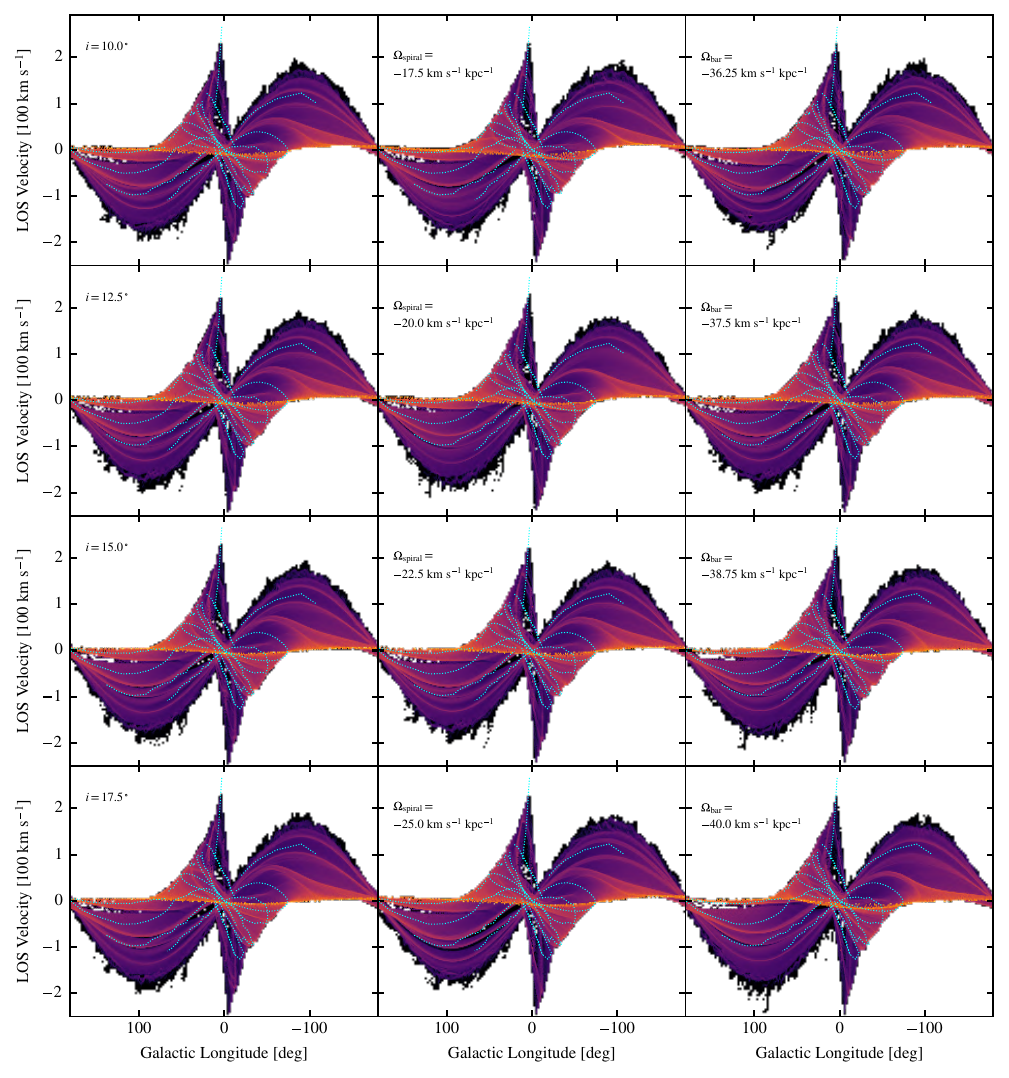}
    \caption{Longitude velocity diagrams of our parameter study with $n=2$ spiral arms with similar phase angle between the spiral arm potential and the bar. Left column is where we vary the pitch angle. The middle and right columns are the variation in spiral arm pattern speed and bar pattern speed respectively.}
    \label{fig:2_spiral_arm_param}
\end{figure*}

\begin{figure*}
    \centering
    \includegraphics{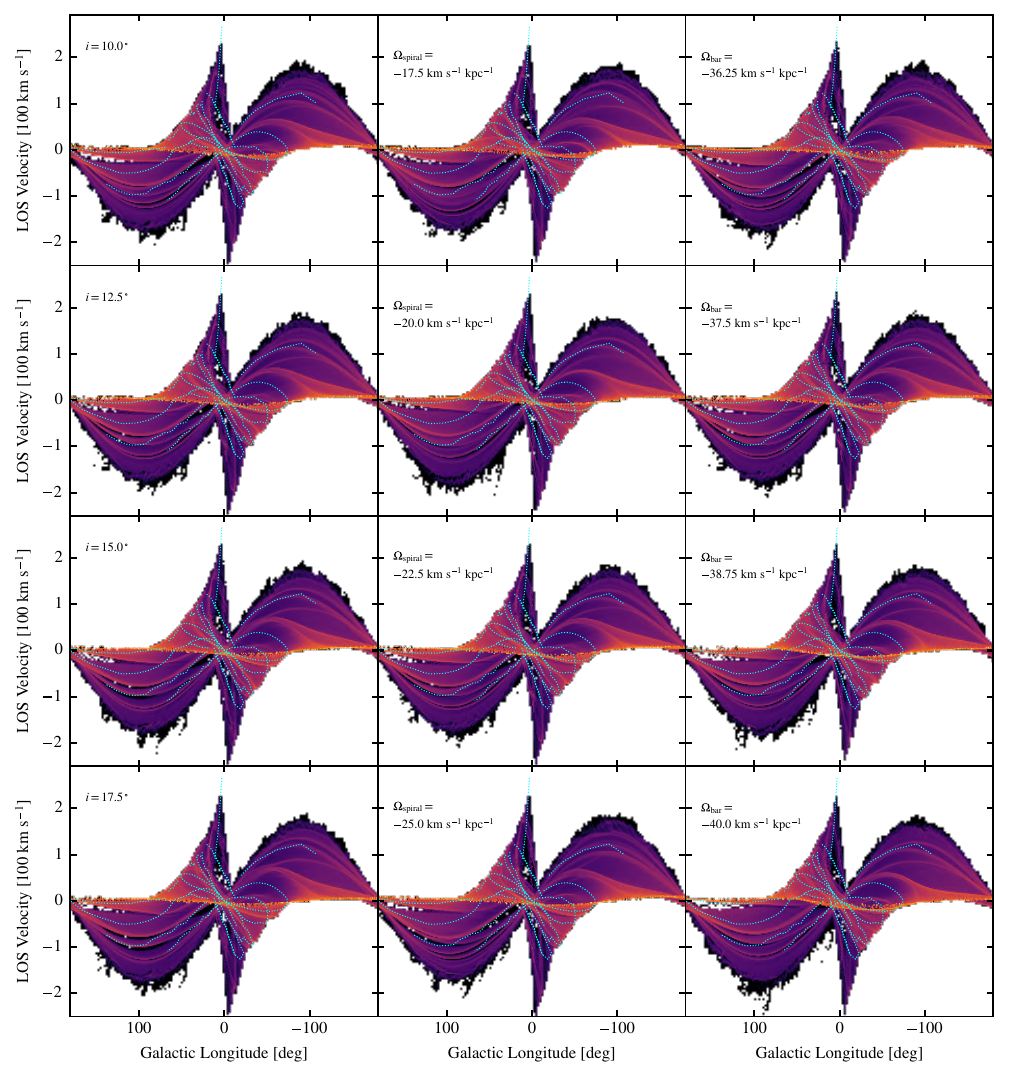}
    \caption{Same as Figure~\ref{fig:2_spiral_arm_param} but with the $n=4$ spiral arms subset.}
    \label{fig:4_spiral_arn_param}
\end{figure*}

We present the $lv$ diagrams of our parameter study in Figures~\ref{fig:2_spiral_arm_param} \& \ref{fig:4_spiral_arn_param}. The first thing that stands out between the figures is the number of filamentary structures in the $lv$ diagram away from the centre of the galaxy, most notably the feature that goes between two spiral arms in the region of $\sim-50^\circ$ to $\sim-130^\circ$ that is present in the $n=4$ set but not in the $n=2$ set. At positive galactic longitudes, we find the expected additional spiral arms with the $n=4$ set. We find that two of the arms in this region of the $lv$ diagram lie very close to each other for the $n=4$, appearing to have split from a spiral arm in the same region in $lv$ space in the $n=2$ set.

From the first column of both figures we see that pitch angle has little impact on the shape of the spiral arms in $lv$ space outside the central region. However, within the central $60^\circ$, differences can be seen between $lv$ diagrams. Here some features move towards the Galactic centre in $lv$ space with increasing pitch angle though not all features, with those associated with the bar's rotation remaining fixed in position.

The spiral arm pattern speed has a larger impact as the corresponding resonances for the spiral arms end up changing with pattern speed. This in turn causes the spiral arms to become more apparent with increasing pattern speed beyond the central region of the $lv$ diagram. Additionally, the features associated with the spiral arms tend towards 0 km s$^{-1}$ in $lv$ space. Within the inner most $60^\circ$ of the Galaxy, the structure here also moves similarly to how it does with pitch angle, moving towards the Galactic \respA{centre} with increasing pattern speed, however the features here that move are different to those with increasing pitch angle suggesting these are resonance features from the spiral arms. 

Similar to the spiral arm pattern speed, changes to bar pattern speed gives arise to different positions for the resonances of the bar, with them moving inward towards $R=0$ kpc with increasing pattern speed (see Figure~\ref{fig:epicyc}). In the $lv$ diagram we see an effect similar to that of the spiral arm pattern speed. Here we see the spiral arms generated by the bar moving towards 0 km s$^{-1}$ in $lv$ space, however the broadening of the features does not happen in this case.

It should be noted for all $lv$ diagrams presented in this section have been selected to have approximately the same phase angle between the bar and spiral arms. This means that whilst each snapshot will has similar phase angle between bar and spiral arms, they will be at different stages in evolution. We select snapshots as close as possible in time late into the systems' evolution ($t>381$ Myr), however there is a range of 147 Myr within the snapshots selected. 

In all cases, the resulting $lv$ diagrams are similar enough that altering one parameter within the constraints of our parameter range does not induce large deviations. However, an exact match to the spiral arm models of \citet{McClure-Griffiths2004}, \citet{Reid2014} and \citet{Reid2019} requires more sophisticated modeling that is beyond the scope of this paper.

\begin{table}
    \caption{Parameter values considered.}
    \begin{tabular}{lcc}
        $\!\!$Parameter$\!\!\!\!\!\!\!\!\!\!\!\!\!\!$ &  Values & Unit \\\hline
        $\!\!$$n$ & 2, 4 &  \\
        $\!\!$$i$ & 10, 12.5$^*$, 15, 17.5 & [$^\circ$]\\
        $\!\!$$\Omega_\mathrm{spiral}$ & $-17.5$, $-20$, $-22.5^*$, $-25$ & [km s$^{-1}$ kpc$^{-1}$] \\
        $\!\!$$\Omega_\mathrm{bar}$ & $-36.25$, $-37.5^*$, $-38.75$, $-40$ & [km s$^{-1}$ kpc$^{-1}$]\\ \hline
    \end{tabular}
    \label{tab:param_study}
    \tablefoot{ Parameters as defined in the main text, with $^*$ indicating fiducial values.}
\end{table}

\end{appendix}


\label{lastpage}
\end{document}